\documentclass[aps,prl,twocolumn,groupedaddress]{revtex4}
\usepackage{amsfonts}
\usepackage{amssymb}
\usepackage{amsmath}

\input{tcilatex}

\begin{document}

\title{Variational principle for the Wheeler-Feynman electrodynamics}
\author{Jayme De Luca}
\email[author's email address:]{ deluca@df.ufscar.br}
\affiliation{Universidade Federal de S\~{a}o Carlos, \\
Departamento de F\'{\i}sica\\
Rodovia Washington Luis, km 235\\
Caixa Postal 676, S\~{a}o Carlos, S\~{a}o Paulo 13565-905\\
Brazil}
\date{\today }

\begin{abstract}
We adapt the formally-defined Fokker action into a variational principle for
the electromagnetic two-body problem. We introduce properly defined boundary
conditions to construct a Poincar\`{e}-invariant-action-functional of a
finite orbital segment\emph{\ }into the reals. The boundary conditions for
the variational principle are an endpoint along each trajectory plus the
respective segment of trajectory for the other particle inside the lightcone
of each endpoint. We show that the conditions for an extremum of our
functional are the mixed-type-neutral-equations with implicit
state-dependent-delay of the electromagnetic-two-body problem. We put the
functional on a natural Banach space and show that the functional is Frech%
\'{e}t-differentiable. We develop a method to calculate the second variation
for $C^{2}$ orbital perturbations in general and in particular about
circular orbits of large enough radii. We prove that our functional has a
local minimum at circular orbits of large enough radii, at variance with the
limiting Kepler action that has a minimum at circular orbits of arbitrary
radii. Our results suggest a bifurcation at some $O(1)$ radius below which
the circular orbits become saddle-point extrema. We give a precise
definition for the distributional-like integrals of the Fokker action and
discuss a generalization to a Sobolev space $H_{0}^{2}$ of trajectories
where the equations of motion are satisfied almost everywhere. Last, we
discuss the existence of solutions for the state-dependent delay equations
with slightly perturbated arcs of circle as the boundary conditions and the
possibility of nontrivial solenoidal orbits.
\end{abstract}

\maketitle

\section{\protect\bigskip Introduction}

We construct a variational principle for the electromagnetic two-body
problem with\emph{\ finite} integration limits. Unlike the Fokker action
that involves an infinite integration and has a formal meaning only\cite%
{Fey-Whe,Schwarz, Tetrode-Fokker}, our Poincar\'{e}-invariant functional
maps a \emph{finite segment }of trajectory into a \emph{finite} real number.
Our variational principle uses boundary conditions consisting of (i) the
initial point $O_{A}$ for the trajectory of particle $1$ plus the segment of
trajectory of particle $2$ inside the lightcone of $O_{A}$, and \ (ii) the
endpoint $L_{B}$ for the trajectory of particle $2$ plus the segment of
trajectory of particle $1$ inside the lightcone of $L_{B}$. For trajectories
respecting the above boundaries we show that the conditions for an extremum
of our functional are the two-body equations of motion of the
Wheeler-Feynman electrodynamics\cite{Fey-Whe}. Our first functional is the
natural generalization of the Fokker action and it can not be defined for
trajectories travelling faster than light (superluminal). We construct a
norm for the linear space of $C^{1}$ orbits satisfying the above boundaries
and show that our functional is Frech\'{e}t differentiable at subluminal
orbits. In order to obtain a functional defined everywhere on a natural
Banach space of $C^{1}$ orbits we give up the parametrization-independence
and construct a second generalized functional. The electromagnetic equations
of motion follow from the conditions for an extremum of our functionals in
the subspace of $C^{2}$ orbital variations. The extremum conditions are
parametrization-independent for the Fokker-like functional, while for the
generalized functional the conditions require the parameter to be
proper-time because of a conservation law that separates the extremal orbits
in three classes. The domain of our second functional is a Banach space, and
moreover along its extremal orbits the particle trajectories turn out to be
of three possible types (a) subluminal trajectories travelling slower than
light everywhere, (b) luminal trajectories travelling \emph{at} the speed of
light everywhere or (c) superluminal trajectories travelling faster than
light everywhere. We calculate the second variation of the action about
general orbits for $C^{2}$ orbital variations satisfying the above boundary
conditions and in particular about the
Schoenberg-Schild-circular-orbit-solutions of a large radius\cite%
{Schoenberg,Schild}. We prove that the second variation is positive-definite
about circular orbits of large enough radii, so that circular orbits are
local minima of our functionals. Our results suggest a bifurcation at some $%
O(1)$ radius below which circular orbits become saddle-point extrema, at
variance with the Kepler action for which circular orbits of arbitrary radii
are minima\cite{Gordon}. We discuss a use of the variational principle to
solve the neutral-delay equations of the electromagnetic two-body problem as
a boundary-value problem with a variational integrator \cite{Marsden}. We
discuss a generalization to a Sobolev space $H_{0}^{2}$ of trajectories
where the equations of motion are valid almost everywhere and the existence
of solutions with slightly perturbed circular boundaries. Last, we discuss
the physics of the Fokker action and the existence of nontrivial solenoidal
orbits.

The Fokker action functional is a synthetic principle of electrodynamics
discovered in the early 20th century\cite{Schwarz, Tetrode-Fokker} and used
in 1945 by Wheeler and Feynman\cite{Fey-Whe} to construct an electrodynamics
of point charges. The Wheeler-Feynman electrodynamics is an alternative
description of classical electromagnetism that avoids the notion of field to
describe the classical laws of Gauss, Faraday, Amp\`{e}re, and Biot-Savart 
\cite{Fey-Whe,Leiter}. The theory describes point charges interacting in
pairs via the half-retarded plus half-advanced solutions of Maxwell's
equations for the fields\cite{Jackson}. Here we avoid the popular name 
\textit{action-at-a-distance electrodynamics} because it can suggest
action-at-the-same-time connecting spatially-separated points, while the
Wheeler-Feynman theory involves only Einstein-local interactions along
lightcones. Among the existing versions of electrodynamics of point charges%
\cite{EliezerReview}, the selling points of the Wheeler-Feynman theory are
(i) The point-charge-limit is regular, i.e., a spherical charge distribution
of a small radius does not make a force on itself and its mass is not
renormalized and (ii) The theory reduces to the usual Dirac electrodynamics
with retarded-only interactions\cite{Dirac} when the far fields vanish
asymptotically, a condition named \textit{the absorber hypothesis} in Refs. 
\cite{Fey-Whe}. The equations for two-body motion of the Wheeler-Feynman
theory are state-dependent neutral-delay equations and little is known about
their solutions, besides the existence of a one-parameter family of
circular-orbit solutions\cite{Schoenberg,Schild}. An existence result was
proved in Ref. \cite{Driverfuture} for the two-body problem with equal
charges (repulsive interaction) and initial condition restricted to colinear
orbits of large separations, a case where the equations of motion are no
longer neutral but rather delay-only. References \cite%
{neutralDriver,MixedDriver} considered satisfying the state-dependent
neutral-delay equations almost-everywhere. In Ref. \cite{JMP} the equations
of motion were expressed as an algebraic-differential system by solving for
the most advanced accelerations, an approach also used in Ref. \cite%
{wellposed} to prove wellposedness and existence for $C^{\infty \text{ }}$%
initial data consisting of maximal independent past segments. The initial
conditions consisting of maximal independent history segments developed in
Ref. \cite{wellposed} are different from the initial conditions used in our
variational method, which combine future and past data. Last, the simpler
delay-only state-dependent two-body equations with initial condition
restricted to colinear orbits were studied numerically in Ref. \cite{Efy1}
for the case of repulsive interaction and in Ref. \cite{Efy2} for the case
of attractive interaction (opposite charges). This paper aims to introduce
the problem for a mathematical audience. In the introduction we start from
the naive and formal language of physics, posing the problem first at an
intuitive level. In the later sections we make an attempt to proceed with
rigor and precise definitions by presenting the results in the form of
theorems.

The paper is divided as follows: In section $1$ we give a crash review of
Minkowski spaces and put in one place the ingredients later used to show
that the equations of motion separate the orbits in three invariant classes
and to construct an action defined on a Banach space. In section $2$ we
introduce the finite action and the boundary conditions. We construct a norm
enforcing the property that perturbations with a small norm of subluminal
orbits yield subluminal orbits. For such norm the functional is Frech\'{e}%
t-differentiable along subluminal orbits. In order to obtain a functional
defined on a complete normed space we relax the parametrization independence
requirement and construct a second functional that can be extended to all
types of orbits of a natural Banach space. In this section we discuss the
advantages of using the variational method as an alternative to the
state-dependent neutral-delay equations of motion as far as numerical
stability. In section $3$ we give a method to calculate the second variation
about arbitrary solutions for $C^{2}$ orbital perturbations. In particular
we calculate the second variation about
low-velocity-circular-orbit-solutions. We show that the quadratic form of
second variation about low-velocity-circular-orbits is positive-definite if
the circular radius is large enough. In this section we develop the idea of
a sewing grid which appears naturally in the integration of the quadratic
form with delay and is useful for the numerical analysis of state-dependent
delay problems. In section $4$ we put the discussions and the conclusion. We
discuss the variational method as a tool to investigate solenoidal and other
types of orbits of the electromagnetic two-body problem. We also discuss the
variational problem with slightly perturbed circular-orbit boundary data.
Last, in the appendix we review the physics of the Fokker action and the
conserved momenta of Noether's theorem. We discuss the nontrivial
possibility of solenoidal orbits with both particles gyrating near the speed
of light with finite and small momenta.

\section{ Preliminaries and Definitions}

We start by explaining the natural coordinatization for Lorentz-invariant
dynamics, i.e., the Lorentz four-space $\tciLaplace ^{4}$ attached to an
inertial frame by Einstein synchronization of clocks (the $\tciLaplace $ in $%
\tciLaplace ^{4}$ stands for Lorentz). A point in $\tciLaplace ^{4}$ is
defined by a time $t$\ and a spatial position $\vec{r}$ \ in the inertial
frame, $\boldsymbol{x}^{\mu }\equiv (t,\vec{r})$, henceforth called the time
component $t$ and the three-vector spatial component $\vec{r}$. The index $%
\mu $ belongs to $(1,2,3,4)$, with $\mu =1$ denoting the time-component and $%
\mu =2,3,4$ denoting the spatial components. From any Minkowski vector $%
\boldsymbol{a}^{\mu }=(a_{o},\vec{a})$ we define its dual vector by$%
\boldsymbol{a}_{\mu }\equiv (a_{o},-\vec{a})$. \ The Minkowski scalar
product is a bilinear product defined as the usual scalar product on $%
\mathbb{R}
^{4}$ between the first vector and the second vector's dual (or vice-versa),
i.e., $(\boldsymbol{a}\cdot \boldsymbol{b})\equiv
a_{1}b_{1}-a_{2}b_{2}-a_{3}b_{3}-a_{4}b_{4}$\ . This definition gives only a
pseudo-scalar bilinear product, and the pseudo-norm $|\boldsymbol{a}%
|^{2}\equiv (\boldsymbol{a}\cdot \boldsymbol{a})$ induced by the Minkowski
product is sensible only for time-like vectors, i.e., when $(\boldsymbol{a}%
\cdot \boldsymbol{a})\geq 0$. The Minkowski product divides the vectors of $%
\tciLaplace ^{4}$ in three classes : (i) if $(\boldsymbol{a}\cdot 
\boldsymbol{a})>0$ the vector is called time-like (for example the
four-velocity along a subluminal orbit), (ii) if $(\boldsymbol{a}\cdot 
\boldsymbol{a})<0$ \ the vector is called space-like (for example the
four-acceleration of a subluminal orbit) and last (iii) if $(\boldsymbol{a}%
\cdot \boldsymbol{a})=0$ the vector is called a null-vector or light-like.
The four-vectors $\boldsymbol{a}$ and $\boldsymbol{b}$ are said to be
orthogonal if $(\boldsymbol{a}\cdot \boldsymbol{b})=0$. The properties of
the Lorentz group and the Minkowski product are discussed in Ref. \cite%
{Barut}, of which we list a few:--(a)\ Two orthogonal light-like vectors are
necessarily multiples of each other because $(\boldsymbol{a}\cdot 
\boldsymbol{a})=(\boldsymbol{b}\cdot \boldsymbol{b})=(\boldsymbol{a}\cdot 
\boldsymbol{b})=0$ implies the Cartesian product of the three-vector
components satisfies $\vec{r}_{a}\cdot \vec{r}_{a}=||\vec{r}_{a}||||\vec{r}%
_{b}||$ (double bars denoting the Euclidean modulus of the three-vector),
(b) All vectors orthogonal to a time-like vector are space-like and form a
three-dimensional space-like subspace. (c) Given a time-like four-vector $%
\boldsymbol{z}$ and an arbitrary four-vector $\boldsymbol{x}$ there is a
unique decomposition $\boldsymbol{x}=\boldsymbol{y}+\alpha \boldsymbol{z}$,
where $\boldsymbol{y}$ is space-like and $\alpha $ a real scalar, and (d)
any orthogonal basis for$\tciLaplace ^{4}$ must contain one time-like
four-vector and 3 space-like four-vectors\cite{Barut}. (e) For time-like
vectors the invariant reverse-Schwartz-inequality holds for the Minkowski
product, i.e., $(\boldsymbol{a}\cdot \boldsymbol{b})^{2}\geq (\boldsymbol{a}%
\cdot \boldsymbol{a})(\boldsymbol{b}\cdot \boldsymbol{b})$, equality holding 
\emph{iff} the vectors are parallel, and last (f) For a time-like and a
space-like vectors, the reverse Schwartz holds without the equal-sign case,
i.e., $(\boldsymbol{a}\cdot \boldsymbol{b})^{2}>(\boldsymbol{a}\cdot 
\boldsymbol{a})(\boldsymbol{b}\cdot \boldsymbol{b})$ \cite{Barut}. The
positivity of the Minkowski product $(\boldsymbol{a}\cdot \boldsymbol{a})$ 
\emph{\ }for the four-velocity in arbitrary parametrization is the physical
condition that the particle travels slower than light. The four-velocity is
light-like in the limit situation when the particle travels \emph{at} the
speed of light. The Minkowski scalar product is left invariant by Lorentz
transformations and it is useful to express the equations of motion and the
action functional in a form explicitly equivariant under the Lorentz group.
The last three components of a Minkowski vector form a spatial three-vector
usually treated differently from the first component, and the various norms
used in this papers are henceforth denoted as follows: (I) the Minkowski
norm is indicated with single bars, i.e., $|\boldsymbol{a}|$, (II) the
Euclidean $%
\mathbb{R}
^{3}$norm of the spatial three-vector part is indicated by double bars,
i.e., $||\vec{r}_{a}||$. We also use double bars to indicate the absolute
value of a real number and (III) the Euclidean $%
\mathbb{R}
^{4}$ norm of a four-vector is indicated by double bars with sub-index $4$,
i.e., $||\boldsymbol{a}||_{4}$ and last (IV) The norm defined on our
functional linear space of trajectories of section 3 is denoted by $|%
\boldsymbol{b}_{1}|_{N(x_{1})}$.

To abbreviate the notation, we henceforth drop the 4-index label and keep
only a lower index $j\in (1,2)$ to identify each particle of the two-body
problem, e.g., $j=1$ denotes electronic four-vector quantities and $\ j=2$
denotes protonic four-vector quantities. For subluminal orbits it is
convenient to express the equations of motion in terms of a
Lorentz-invariant parameter defined by the squared-Minkowski-norm of the
infinitesimal displacement vector $d\boldsymbol{x}_{i}$ , i.e., 
\begin{equation}
(d\tau _{i})^{2}=(dt_{i})^{2}-(dx_{i})^{2}-(dy_{i})^{2}-(dz_{i})^{2}>0.
\label{dtau}
\end{equation}%
The left-hand side of Eq. (\ref{dtau}) is positive for subluminal orbits,
zero for luminal orbits and negative for superluminal orbits. The parameter $%
\tau _{i}$ defined by Eq. (\ref{dtau}) is called the proper-time and it is a
property of each particle's trajectory, the usual parametrization by
arc-length of differential geometry.

Next we introduce the naive Fokker action in the above defined Lorentz
four-space $\tciLaplace ^{4}$ using a normalized unit system where the speed
of light is $c\equiv 1$ and the electron and the proton have mass and charge 
$m_{1}=1$ and $e_{1}=-1$ and $m_{2}=1824$ and $e_{2}=1$ respectively. Let
the trajectory of each particle in $\tciLaplace ^{4}$ be a differentiable
function $\boldsymbol{x}_{i}(\lambda _{i})$ $:$ $%
\mathbb{R}
\rightarrow \tciLaplace ^{4}$ of a parameter $\lambda _{i}$ with $i=1,2$
indicating respectively the electron and the proton trajectories. The Fokker
action\cite{Schwarz, Tetrode-Fokker} is defined in the original literature
by a \emph{formal }integration along the whole trajectories as 
\begin{eqnarray}
S &=&-\int m_{1}\sqrt{\boldsymbol{\dot{x}}_{1}\cdot \boldsymbol{\dot{x}}_{1}}%
d\lambda _{1}-\int m_{2}\sqrt{\boldsymbol{\dot{x}}_{2}\cdot \boldsymbol{\dot{%
x}}_{2}}d\lambda _{2}  \notag \\
&&+\int \int \delta (|\boldsymbol{x}_{1}-\boldsymbol{x}_{2}|^{2})\boldsymbol{%
\dot{x}}_{1}\cdot \boldsymbol{\dot{x}}_{2}d\lambda _{1}d\lambda _{2},
\label{aFokker}
\end{eqnarray}%
where overdot denotes derivative respect to the parameter of each
trajectory. Action (\ref{aFokker}) is formally independent of the
parametrizations, a geometric property easily checked by changing the
parameter of each trajectory with the chain rule. The peculiar last integral
of the right-hand-side of Eq. (\ref{aFokker}) involves the composition of
the Dirac delta-function $\delta (x)$ with the\ real function 
\begin{equation}
d(\lambda _{1,}\lambda _{2})\equiv |\boldsymbol{x}_{1}(\lambda _{1})-%
\boldsymbol{x}_{2}(\lambda _{2})|^{2},  \label{light-cone}
\end{equation}%
where single bars stand for the Minkowski norm of the four-separation $%
\boldsymbol{x}_{12}\equiv (\boldsymbol{x}_{1}-\boldsymbol{x}_{2})$. The
peculiar combination appearing in Eq. (\ref{aFokker}) comes from the Green's
function of Maxwell's equations and becomes ill-defined along $C^{1}$
trajectories or in a Sobolev space. Here we give a precise definition for
the right-hand-side of Eq. (\ref{aFokker}), and in Section 3 we define a
consistent derivative for such operation before evaluating the second
variation, thus avoiding the loose derivatives of the Dirac delta-function.
Condition (\ref{dtau}) is sufficient for the separation $d(\lambda
_{1},\lambda _{2})$ of Eq. (\ref{light-cone}) to have precisely two zeros
for each fixed $\lambda _{1}$ along a $C^{1}$ trajectory \cite{wellposed}.
In Ref. \cite{wellposed} it is proved that along a subluminal orbit light
captures the slower moving particle once in the past and once in the future.
The integration on the last term of the right-hand-side of Eq. (\ref{aFokker}%
) gives a nonzero contribution at each zero $(\lambda _{1},$ $\lambda _{2})$
of Eq. (\ref{light-cone}). At a given $\lambda _{1}$ the condition%
\begin{equation}
|\boldsymbol{x}_{1}(\lambda _{1})-\boldsymbol{x}_{2}(\lambda _{2})|^{2}=0,
\label{implicit}
\end{equation}%
can be solved for the time-component $t_{2}(\lambda _{2})$ of vector $%
\boldsymbol{x}_{2}(\lambda _{2})$, yielding a retarded time and an advanced
time, each defined implicitly by 
\begin{equation}
t_{2}(\lambda _{2})=t_{1}(\lambda _{1})\mp ||\vec{r}_{2}(\lambda _{2})-\vec{r%
}_{1}(\lambda _{1})||.  \label{retarded light-cone}
\end{equation}%
where double bars stand for the Euclidean norm of the spatial separation $%
\vec{r}_{2}(\lambda _{2})-\vec{r}_{1}(\lambda _{1})$. Either one of the
equivalent Eqs. (\ref{implicit}) or (\ref{retarded light-cone}) are
henceforth called the lightcone condition. Equation (\ref{retarded
light-cone}) is an implicit condition for $\lambda _{2}$ because $\lambda
_{2}$ appears on both sides as an unknown argument. Condition (\ref{implicit}%
) is symmetric on particle trajectories, so that the lightcone condition for
the protonic trajectory is still Eq. (\ref{retarded light-cone}), as
obtained by either rearranging Eq. (\ref{retarded light-cone}) to isolate $%
t_{1}$ on the left-hand-side or by exchanging the indices $1$ and $2$ of Eq.
(\ref{retarded light-cone}). In the following we assume the orbital
parameters are restricted to the intervals $[L_{\alpha I},L_{\alpha F}]$ for 
$\alpha =1,2$, as defined in the next section. Let the zeros of $d(\lambda
_{1},\lambda _{2})$ for$\ \lambda _{2}\in $ $[L_{2I},L_{2F}]$ and a fixed $%
\lambda _{1}$ $\in $ $[L_{1I},L_{1F}]$ be $(\lambda _{1},\lambda
_{2}^{(j)}(\lambda _{1}))$. \textit{Definition 1}: We henceforth \emph{define%
} the integral involving the Dirac delta-function composed with $d(\lambda
_{1},\lambda _{2})$ by%
\begin{equation}
\dint\limits_{L_{2I}}^{L_{2F}}\delta (d(\lambda _{1},\lambda ))f(\lambda
_{1},\lambda )d\lambda \equiv \tsum\limits_{j}\frac{f(\lambda _{1},\lambda
_{2}^{(j)})}{||\frac{\partial d}{\partial \lambda _{2}}(\lambda _{1},\lambda
_{2}^{(j)})||},  \label{limit}
\end{equation}%
where $||\frac{\partial d}{\partial \lambda _{2}}(\lambda _{1},\lambda
_{2}^{(j)})||$ is the absolute value of the partial derivative of $d(\lambda
_{1},\lambda _{2})$ evaluated at each zero $(\lambda _{1},\lambda
_{2}^{(j)}) $ of $d(\lambda _{1},\lambda _{2})$. The summation on the
right-hand-side of Eq. (\ref{limit}) includes all the zeros of the lightcone
condition inside $[L_{2I},L_{2F}]$. Once the separation $d(\lambda
_{1},\lambda _{2})$ is completely symmetric on particle quantities,
definition (\ref{limit}) has a symmetric definition as follows;-- \textit{%
Definition} 2: We henceforth define the integral over $\lambda _{1}$ $\in $ $%
[L_{1I},L_{1F}]$ involving the Dirac delta-function composed with $d(\lambda
_{1},\lambda _{2})$ by 
\begin{equation}
\dint\limits_{L_{1I}}^{L_{1F}}\delta (d(\lambda ,\lambda _{2}))f(\lambda
,\lambda _{2})d\lambda \equiv \tsum\limits_{k}\frac{f(\lambda
_{1}^{(k)},\lambda _{2})}{||\frac{\partial d}{\partial \lambda _{1}}(\lambda
_{1}^{(k)},\lambda _{2})||},  \label{limit1}
\end{equation}%
where $||\frac{\partial d}{\partial \lambda _{1}}(\lambda _{1}^{(k)},\lambda
_{2})||$ is the absolute value of the partial derivative of $d(\lambda
_{1},\lambda _{2})$ evaluated at each zero $(\lambda _{1}^{(k)},\lambda
_{2}) $ of $d(\lambda _{1},\lambda _{2})$ for a fixed $\lambda _{2}$ $\in $ $%
[L_{2I},L_{2F}]$ and $\lambda _{1}\in $ $[L_{1I},L_{1F}]$. For subluminal
orbits the interval $[L_{kI},L_{kF}]$ can include at the most the two zeros
proved in Ref. \cite{wellposed}, while for superluminal orbits there can be
several zeros inside $[L_{kI},L_{kF}]$, or even none. If no zero exists in
the integration interval the right-hand side of either Eqs. (\ref{limit}) or
(\ref{limit1}) is defined to be \emph{zero}. Definitions (\ref{limit}) and (%
\ref{limit1}) are motivated by the evaluation of the respective left-hand
sides of Eqs. (\ref{limit}) and (\ref{limit1}) using the Dirac
delta-function with a $C^{\infty }$ separation $d(\lambda _{1},\lambda _{2})$
and changing variables using the absolute value of the Jacobian of the local
coordinate change near each zero. Here we avoid distributional operations
with the Dirac delta-function and henceforth take Eqs. (\ref{limit}) and (%
\ref{limit1}) as defining a functional of $d(\lambda _{1},\lambda _{2})$, $%
f(\lambda _{1},\lambda _{2})$ and the intervals $[L_{kI},L_{kF}]$ into the
reals. It is further useful to define the function $A(\boldsymbol{%
\boldsymbol{x}_{j})}:\tciLaplace ^{4}\rightarrow \tciLaplace ^{4}$ by 
\begin{equation}
\boldsymbol{A}_{k}\boldsymbol{(\boldsymbol{x}_{j}(\lambda }_{j}\boldsymbol{%
))\equiv }\dint\limits_{L_{kI}}^{L_{kF}}\delta (|\boldsymbol{x}_{k}(\lambda
_{k})-\boldsymbol{x}_{j}(\lambda _{j})|^{2})\boldsymbol{\dot{x}}_{k}(\lambda
_{k})d\lambda _{k},  \label{defvecA}
\end{equation}%
where $(k,j)=(1,2)$ or $(2,1)$ and the integration on the right-hand side of
(\ref{defvecA}) is \textit{defined }either by Eq. (\ref{limit}) or Eq. (\ref%
{limit1}). The vector function defined by Eq. (\ref{defvecA}) is often
called the vector-potential in physics. Assuming integral (\ref{defvecA}) to
exist for both $(k,j)=(1,2)$ and $(2,1)$, the interaction double-integral of
the right-hand-side of Eq. (\ref{aFokker}) can be expressed by%
\begin{equation}
I=\tint\limits_{L_{jI}}^{L_{jF}}\boldsymbol{A}_{k}\boldsymbol{(\boldsymbol{x}%
_{j})\cdot \dot{x}}_{j}d\lambda _{j},  \label{defI}
\end{equation}%
with either $(k,j)=(1,2)$ or $(2,1)$. Using either definition (\ref{limit})
or (\ref{limit1}) we can express the interaction term (\ref{defI}) in the
two equivalent forms 
\begin{eqnarray}
I &=&\tint\limits_{L_{2I}}^{L_{2F}}d\lambda _{2}\tsum\limits_{k}\frac{%
\boldsymbol{\dot{x}}_{1}(\lambda _{1}^{(k)})\cdot \boldsymbol{\dot{x}}%
_{2}(\lambda _{2})}{||\frac{\partial d}{\partial \lambda _{1}}(\lambda
_{1}^{(k)},\lambda _{2})||}  \label{equivalent} \\
&=&\tint\limits_{L_{1I}}^{L_{1F}}d\lambda _{1}\tsum\limits_{j}\frac{%
\boldsymbol{\dot{x}}_{1}(\lambda _{1})\cdot \boldsymbol{\dot{x}}_{2}(\lambda
_{2}^{(j)})}{||\frac{\partial d}{\partial \lambda _{2}}(\lambda _{1},\lambda
_{2}^{(j)})||}.  \notag
\end{eqnarray}%
It is instructive to check the equivalence of formulas (\ref{equivalent}) by
changing the integration variable from $\lambda _{2}$ to $\lambda _{1}$
about each zero. Condition (\ref{implicit}) defines $\lambda _{1}$ as a
function of $\lambda _{2}$ by the the implicit function theorem and the
Jacobian of the coordinate change transforms the first line of Eq.(\ref%
{equivalent}) into the second line of \bigskip Eq.(\ref{equivalent}). Last,
to express the Jacobian in the usual form of physics textbooks we define%
\begin{eqnarray}
J_{\lambda _{2}}^{\pm } &\equiv &-\frac{1}{2}\frac{\partial d}{\partial
\lambda _{2}}(\lambda _{1},\lambda _{2}^{\pm })  \label{Jacobian} \\
&=&[\boldsymbol{x}_{1}(\lambda _{1})-\boldsymbol{x}_{2}(\lambda _{2}^{\pm
})]\cdot \boldsymbol{\dot{x}}_{2}(\lambda _{2}^{\pm }).  \notag
\end{eqnarray}%
For subluminal orbits $\boldsymbol{\dot{x}}_{2}(\lambda _{2}^{\pm })$ is a
time-like vector with a positive time-velocity, and once $\boldsymbol{x}%
_{1}(\lambda _{1})-\boldsymbol{x}_{2}(\lambda _{2}^{\pm })$ is a
null-vector, condition \ref{Jacobian} defines a positive $J_{\lambda
_{2}}^{-}$ on the retarded lightcone and a negative $J_{\lambda _{2}}^{+}$
on the advanced cone. For superluminal orbits $J_{\lambda _{2}}^{\pm }$ can
have any sign in either lightcone, so that it is best to keep the moduli in
the denominators of (\ref{limit}).

\section{Action with bounds}

The guiding principle to construct an action functional is that the extremum
condition should generate the electromagnetic equations of motion\cite%
{Schwarz, Tetrode-Fokker,Fey-Whe}. In the following we start from the naive
Fokker action (\ref{aFokker}) and explain how to restrict the integration to
suitable finite segments of trajectory, using particle-time parametrization
just for simplicity of the exposition. The original works \cite{Schwarz,
Tetrode-Fokker,Fey-Whe} extended the integration of (\ref{aFokker}) from
plus to minus infinity as a simple solution to include the needed future or
past of the other particle at endpoints. There is no reason to assume such
integration should converge, so that the infinite integral (\ref{aFokker})
has a formal-only meaning\cite{StarusCritique}. Moreover, the Fokker action
yields the electromagnetic equations of motion\cite{Schwarz,
Tetrode-Fokker,Fey-Whe}\emph{\ only }if the condition of extremum is
enforced \emph{formally} with trajectory variations of compact support. Here
we avoid the shortcomings of a formal-only action and give instead a
finite-valued functional. The boundary conditions can be restricted to a
point and a segment along each trajectory in a way that the future and the
past lightcone points exist everywhere along both trajectories, as follows;
\ Let the initial point of trajectory $1$ be point $O_{A}$ at $t_{1}=0$ and
the endpoint of trajectory $2$ be point $L_{B}$ at $t_{2}=T_{2}$ as
illustrated in Fig. 3.1. The trajectory of particle $1$ to be varied extends
from $O_{A}$ to point $L^{-}$ at $t_{1}=T_{1}$ where trajectory $1$
intersects the past lightcone of $L_{B}$ ( indicated in\ green in Fig.1).
The future history of particle $1$ is needed from point $L^{-}$ to point $%
L^{+}$ at $t_{1}=\Lambda _{1}^{+}>T_{1}$ where trajectory $1$ intersects the
advanced lightcone of $L_{B}$ (the red portion of the upper trajectory of
Fig. 3.1). The past history of particle $2$ is needed from point $O^{-}$ at $%
t_{2}=\Lambda _{2}^{-}$ where trajectory $2$ intersects the past lightcone
of $O_{A}$ up to point $O^{+}$at $t_{2}=\Lambda _{2}^{+}<T_{2}$ where
trajectory $2$ intersects the future lightcone of $O_{A}$ (also indicated in
red in Fig.1). The trajectory of particle $2$ to be varied goes from $O^{+}$
to $L_{B}$. The combination of the initial point $O_{A}$ along trajectory $1$
and the final point $L_{B}$ along trajectory $2$, plus the respective
segments of trajectory inside the lightcones of these endpoints is
henceforth called exchange-of-history boundary conditions (EHBCs) as
indicated in red in Fig. 3.1. Our construction is Lorentz-invariant because
lightcones are Lorentz-invariant objects. The construction is unique up to a
time-reversed construction using an endpoint along trajectory $1$ and an
initial point along trajectory $2$ plus the history segments inside the
respective lightcones. For $C^{1}$ orbits the EHBCs complete the
trajectories in such a way that any point along each trajectory has the two
lightcone roots inside the evaluation interval for either one of the
interaction formulas (\ref{equivalent}).

The restrictions for the EHBCs histories are; (a) it must be possible to
travel from the initial point to the final point of each trajectory at a
speed lesser (or equal at the most) than light, and (b) The \textit{%
minimally short} condition that trajectory $1$ must intersect the future
lightcone of $O^{+}$ \emph{before} arriving at endpoint $L^{-}$ (at time $%
t_{1}=T_{1}$). In this way the past history of particle $2$ does not
interact with the future history of particle $1$. Beyond that the
variational method can be postulated with otherwise arbitrary histories. The
advantage of solving the state-dependent delay equations using the
variational method with the EHBCs\ is the numerical stability:--Once the
equations of motion are time-reversible the stable and unstable manifolds
exist in pairs, so that tying both ends down with the EHBCs avoids the orbit
to diverge either in the future direction along the unstable manifold or in
the past direction along the stable manifold. Since the maximum spatial
velocity is $c=1$, the spatial position of particle $1$ is bounded by a
sphere of radius $T_{1}$ centered at point $O_{A}$, while the spatial
position of particle $2$ is bounded by a sphere of radius $(T_{2}-\Lambda
_{2}^{+})$ centered at $O^{+}$. Therefore the subluminal trajectories
satisfying the EHBCs are spatially bounded and there are no runaway orbits
satisfying the EHBCs during the optimization. The interaction formula of Eq.
(\ref{equivalent}) needs the position in lightcone along the other
trajectory, which is naturally approximated numerically using the
trapezoidal rule with an integration grid consisting of the union of sewing
chains defined as follows;---(i) A forward sewing chain is a set of
consecutive points in lightcone starting from an arbitrary point on the
boundary segment from $O_{-}$ to $O_{+}$ (as illustrated in Fig. 4.1). The
chain goes up to the corresponding point in future lightcone along
trajectory $1$ and \ back down and up until the last point along the
boundary segment of trajectory $1$ from $L^{-}$ to $L^{+}$ and (ii) A
backward sewing chain is a set of consecutive points in lightcone starting
from any point on the boundary segment from $L^{-}$ and $L^{+}$ of the
trajectory $1$ (as illustrated in Fig. 4.1). The sewing chain proceeds to
the corresponding point \ in past lightcone along trajectory $2$ and \ back
down and up until the last backward point on the boundary segment from $%
O_{-} $ to $O_{+}$. It is important to include in the sewing grid the
forward chain starting from $O_{A}$ and the backward chain starting from $%
L_{B}$ because these chains separate boundary data from orbital data. Notice
that a sewing chain starts from a point along one trajectory and ends with a
point along the other trajectory, so that each chain defines the same number
of points along each orbit.

For arbitrary trajectory variations satisfying the EHBCs the linearized
functional variation is a sum of the linear variations along the two special
cases;-- (i) one fixes trajectory $2$ while varying trajectory $1$
arbitrarily and (ii) one fixes trajectory $1$ while varying trajectory $2$
arbitrarily, so that it suffices to study problems (i) and (ii). In the
following we study (i) using particle-time parametrization, for which we
integrate over $t_{2}$ in the double integral of action (\ref{aFokker}) with
the help of Eq. (\ref{limit}). The half-Jacobian needed for Eq. (\ref{limit}%
) is a case of Eq. (\ref{Jacobian}) with the choice of parameter $\lambda
_{2}=t_{2}$, i.e., 
\begin{equation}
J_{t_{2}}\equiv \lbrack \boldsymbol{x}_{1}(t_{1})-\boldsymbol{x}%
_{2}(t_{2})]\cdot \boldsymbol{\dot{x}}_{2}(t_{2}).  \label{Jacobian_t}
\end{equation}%
The dot over $\boldsymbol{x}_{2}$ in Eq. (\ref{Jacobian_t}) denotes
derivative respect to particle-time. Using (\ref{limit}) to integrate over $%
\lambda _{2}=t_{2}$ inside the double-integral on the right-hand side of Eq.
(\ref{aFokker}) yields%
\begin{eqnarray}
S &=&\int_{0}^{T_{1}}-m_{1}\sqrt{\boldsymbol{\dot{x}}_{1}\cdot \boldsymbol{%
\dot{x}}_{1}}dt_{1}+\int_{\Lambda _{2}^{+}}^{T_{2}}-m_{2}\sqrt{\boldsymbol{%
\dot{x}}_{2}\cdot \boldsymbol{\dot{x}}_{2}}dt_{2}  \notag \\
&&+\int_{0}^{T_{1}}\frac{\boldsymbol{\dot{x}}_{1}\cdot \boldsymbol{\dot{x}}%
_{2+}}{2||J_{t_{2}}^{+}||}dt_{1}+\int_{0}^{\Lambda _{1}^{+}}\frac{%
\boldsymbol{\dot{x}}_{1}\cdot \boldsymbol{\dot{x}}_{2-}}{2||J_{t_{2}}^{-}||}%
dt_{1},  \label{Darw}
\end{eqnarray}%
where the superscripts $\pm $ on $J_{t_{2}}^{\pm }$ indicate evaluation on
the advanced/retarded light-cone of particle $1$, respectively. Notice that
action (\ref{Darw}) is defined only for subluminal and luminal orbits. To
evaluate the functional derivative of $S$ with respect to variations of
trajectory $1$ we can drop the last term of the first line on the right-hand
side of Eq. (\ref{Darw}), which is independent of trajectory $1$. Last, the
integration over the future history of particle $1$ extending from $%
t_{1}=T_{1}$ to $t_{1}=\Lambda _{1}^{+}$\ is left invariant by a variation
of trajectory $1$ respecting the EHBCs, so that we can replace the upper
limit of the last integral on the right-hand side of Eq. (\ref{Darw}) by $%
T_{1}$, yielding an integration over $t_{1}\in \lbrack 0,T_{1}]$ of a
Lagrangian function $L_{1}(\boldsymbol{x}_{1},\boldsymbol{\dot{x}}_{1})$
defined as,

\begin{eqnarray}
S_{1} &\equiv &\dint_{0}^{T_{1}}L_{1}(\boldsymbol{x}_{1},\boldsymbol{\dot{x}}%
_{1})dt_{1}  \label{partialLagrangian} \\
&\equiv &\dint_{0}^{T_{1}}[-m_{1}\sqrt{\boldsymbol{\dot{x}}_{1}\cdot 
\boldsymbol{\dot{x}}_{1}}+\frac{\boldsymbol{\dot{x}}_{1}\cdot \boldsymbol{%
\dot{x}}_{2+}}{2||J_{t_{2}}^{+}||}+\frac{\boldsymbol{\dot{x}}_{1}\cdot 
\boldsymbol{\dot{x}}_{2-}}{2||J_{t_{2}}^{-}||}]dt_{1}.  \notag
\end{eqnarray}%
In Eq. (\ref{partialLagrangian}) the advanced/retarded position and velocity
of particle $2$, indicated by $\pm $, are evaluated with the fixed
trajectory of particle $2$ at the advanced/retarded points defined by the
roots $t_{2}^{\pm }(t_{1})$ of Eq. (\ref{implicit}), which are implicit
functions of the updated trajectory of particle $1$. Notice that even though
the trajectory of particle $2$ is fixed, the corresponding lightcone points
move away from $t_{2}^{\pm }$ along the fixed trajectory $2$ as we vary the
trajectory $1$. The gradient of $t_{2}^{\pm }$ respect to the four-position $%
\boldsymbol{x}_{1}$ is obtained relating the differential $d\boldsymbol{x}%
_{1}$ along trajectory $1$ to the differential $dt_{2}^{\pm }$ via the
derivative of the implicit condition (\ref{implicit}), i.e.,%
\begin{equation}
-2J_{t_{2}}^{\pm }dt_{2}^{\pm }+2\boldsymbol{x}_{12}^{\pm }\cdot d%
\boldsymbol{x}_{1}=0,  \label{varia}
\end{equation}%
where $J_{t_{2}}^{\pm }$is defined by Eq. (\ref{Jacobian_t}) and $%
\boldsymbol{x}_{12}^{\pm }\equiv \boldsymbol{x}_{1}(t_{1})-\boldsymbol{x}%
_{2}(t_{2}^{\pm })$. \ Therefore the derivative of $t_{2}^{\pm }(\boldsymbol{%
x}_{1})$ respect to $\mathbf{x}_{1}$ along the fixed orbit of particle $2$
is 
\begin{equation}
\frac{\partial t_{2}^{\pm }}{\partial \boldsymbol{x}_{1}}=\frac{\boldsymbol{x%
}_{12}^{\pm }}{J_{t_{2}}^{\pm }}.  \label{slide}
\end{equation}

Next we construct a linear space consisting of the $C^{1}$ orbital
neighborhood of any $C^{1}$ subluminal orbit $\boldsymbol{x}_{1}$ satisfying
the EHBCs. Without loss of generality we operate with perturbations of
trajectory $1$ only, defined as type (i) in the paragraph above Eq. (\ref%
{Jacobian_t}). \textit{Definition 3}: For a $C^{1}$ subluminal orbit $%
\boldsymbol{x}_{1}$ satisfying the EHBCs we define the linear space $N^{(1)}(%
\boldsymbol{x}_{1})$ as the set of all $C^{1}$ trajectories defined by a
perturbation function $\boldsymbol{b}_{1}\boldsymbol{:}$ $[0,\lambda
_{1F}]\rightarrow \tciLaplace ^{4}$ , i.e., 
\begin{eqnarray}
\boldsymbol{u}_{1} &\equiv &\boldsymbol{x}_{1}+\boldsymbol{b}_{1},
\label{variation} \\
\boldsymbol{\dot{u}}_{1} &\equiv &\boldsymbol{\dot{x}}_{1}+\boldsymbol{\dot{b%
}}_{1},  \notag
\end{eqnarray}%
where $\boldsymbol{b}_{1}$ vanishes at the endpoints in accordance with the
EHBCs, i.e.,

\begin{eqnarray}
\boldsymbol{b}_{1}(\lambda _{1} &=&0)=0,  \label{endpoint} \\
\boldsymbol{b}_{1}(\lambda _{1} &=&\lambda _{1F})=0.  \notag
\end{eqnarray}%
Notice that the EHBCs forbid orbital perturbations $\boldsymbol{b}_{1}%
\boldsymbol{(}\lambda _{1}\boldsymbol{)}$ with a monotonically increasing
time-component\ because condition (\ref{endpoint}) is impossible for a
monotonically increasing time-component. The usual norm for the space of $%
C^{1}$ functions is given by $||\equiv \sup ||\boldsymbol{b}_{1}||_{4}+\sup
||\boldsymbol{\dot{b}}_{1}||_{4}$ and because of conditions (\ref{endpoint})
it turns out that $\sup ||\boldsymbol{b}_{1}||_{4}\leq |\lambda _{1F}|\sup ||%
\boldsymbol{\dot{b}}_{1}||_{4}$ for $0<\lambda _{1}<\lambda _{1F}$ as can be
shown using either one of conditions (\ref{endpoint}). For example using $%
\boldsymbol{b}_{1}(\lambda _{1}=\lambda _{1F})=0$ we have%
\begin{equation}
\boldsymbol{b}_{1}=-\tint\limits_{\lambda _{1}}^{\lambda _{1F}}\boldsymbol{%
\dot{b}}_{1}(\lambda )d\lambda ,  \label{integral}
\end{equation}%
so that $\sup ||\boldsymbol{b}_{1}||_{4}\leq |\lambda _{1F}|\sup ||%
\boldsymbol{\dot{b}}_{1}||_{4}$ for $0<\lambda _{1}<\lambda _{1F}$.
Therefore we can drop the $\sup ||\boldsymbol{b}_{1}||_{4}$ term of the norm
and henceforth our norm is simply defined by the sup of the Euclidean $%
\mathbb{R}
^{4}$ norm of $\boldsymbol{\dot{b}}_{1}$, i.e., $|\boldsymbol{b}%
_{1}|_{N(x_{1})}\equiv \sup ||\boldsymbol{\dot{b}}_{1}||_{4}$. Notice that
whenever $\boldsymbol{\dot{b}}_{1}=0$ the condition $\sup ||\boldsymbol{\dot{%
b}}_{1}||_{4}=0$ plus the endpoint condition (\ref{endpoint}) selects the
single constant element $\boldsymbol{b}_{1}=0$, so that $\sup ||\boldsymbol{%
\dot{b}}_{1}||_{4}$ defines a norm on the linear space of $C^{1}$ functions $%
\boldsymbol{b}_{1}\boldsymbol{:}$ $[0,\lambda _{1F}]\rightarrow \tciLaplace
^{4}$ satisfying the EHBCs. The linear space $N^{(1)}(\boldsymbol{x}_{1})$
can be shown to be a Banach space with this norm in the usual way. \textit{%
Proposition 1} : Subluminal orbits $(t_{1}(\lambda _{1}),\vec{r}_{1}(\lambda
_{1}))$ have small neighborhoods in $N^{(1)}(\boldsymbol{x}_{1})$ containing
only subluminal orbits. To show it we define the local Cartesian velocity
respect to particle-time by%
\begin{equation}
\vec{v}(\lambda _{1})\mathbf{\equiv }\frac{(d\vec{r}_{1}/d\lambda _{1})}{%
(dt_{1}/d\lambda _{1})},  \label{Vt}
\end{equation}%
a three-vector function of $\lambda _{1}$ with Euclidean norm lesser than
one by condition (\ref{dtau}). Along a subluminal orbit condition (\ref{dtau}%
) is positive on the compact set $[0,\lambda _{1F}]$, so that $(1-||\vec{v}%
_{1}||^{2})>\delta >0$ on $[0,\lambda _{1F}]$. \ For any orbit $\boldsymbol{x%
}_{1}(\lambda _{1})$ satisfying Eq.(\ref{dtau}) we can further define $%
h_{1}\equiv (dt_{1}/d\lambda _{1})>0$ and express the velocity $\boldsymbol{%
\dot{x}}_{1}$ by%
\begin{equation}
\boldsymbol{\dot{x}}_{1}=(h_{1},h_{1}\vec{v}_{1}).  \label{physORB}
\end{equation}%
Given a perturbation $\boldsymbol{\dot{b}}_{1}\equiv (\dot{b},\dot{b}\vec{v}%
_{b})$ and a subluminal orbit $\boldsymbol{\dot{x}}_{1}=(h_{1},h_{1}\vec{v}%
_{1})$, substitution of (\ref{variation}) into (\ref{dtau}) yields 
\begin{eqnarray}
\!\!(\boldsymbol{\dot{x}}_{1}+\boldsymbol{\dot{b}}_{1})^{2} &=&(h_{1}+\dot{b}%
)^{2}-||h_{1}\vec{v}_{1}+\dot{b}\vec{v}_{b}||^{2}=  \notag \\
&&h_{1}^{2}(1-||\vec{v}_{1}||^{2})+\dot{b}^{2}(1-||\vec{v}_{b}||^{2})  \notag
\\
&&+2h_{1}\dot{b}(1-\vec{v}_{1}\cdot \vec{v}_{b}),  \label{addition}
\end{eqnarray}%
Since the positivity of Eq. (\ref{addition}) is independent of monotonic
reparametrizations, in the following we use parametrization by the
time-component of $\boldsymbol{x}_{1}$, so that $h_{1}=1$. The norm $|%
\boldsymbol{b}_{1}|_{N(x_{1})}$ dominates the absolute value of the
time-velocity perturbation $\dot{b}$ defined above Eq. (\ref{addition}),
i.e., $|\boldsymbol{b}_{1}|_{N(x_{1})}\geq |\dot{b}|,$ so that one can limit 
$|\dot{b}|$ by choosing $\delta /4>$ $|\boldsymbol{b}_{1}|_{N(x_{1})}\geq |%
\dot{b}|$. \ Equation (\ref{addition}) with $\delta /4>|\dot{b}|$ and $%
h_{1}=1$ shows that the perturbed element is subluminal for small enough $|%
\boldsymbol{b}_{1}|_{N(x_{1})}$, so that subluminal orbits have small
neighborhoods containing only subluminal orbits.

Next we define the Frech\'{e}t derivative of action (\ref{Darw}) about a
subluminal orbit $\boldsymbol{x}_{1}$: Let $S_{1}(\boldsymbol{b}_{1},%
\boldsymbol{\dot{b}}_{1})$ $:$ $N^{(1)}(\boldsymbol{x}_{1})$ $\rightarrow $ $%
\mathbb{R}
$ be defined by substituting (\ref{variation}) into Eq. (\ref%
{partialLagrangian}) and expanding to linear order for small $|\boldsymbol{b}%
_{1}|_{N(x_{1})}$. The linear expansion of $S_{1}(\boldsymbol{b}_{1},%
\boldsymbol{\dot{b}}_{1})$ in terms of $\boldsymbol{b}_{1}$ and $\boldsymbol{%
\dot{b}}_{1}$ involves integrals controlled by an $O(|\boldsymbol{b}%
_{1}|_{N(x_{1})}^{2})$ error in the above defined subliminal neighborhood $|%
\boldsymbol{b}_{1}|_{N(x_{1})}<\delta /4$ because the Euclidean norm $||%
\boldsymbol{b}_{1}||_{4}$ is also bounded by $|\lambda _{1F}|\sup ||%
\boldsymbol{\dot{b}}_{1}||_{4}$ as explained above Eq. (\ref{integral}). \
The linear expansion of $S_{1}$ is already the desired Frech\'{e}t
derivative, i.e.,

\begin{equation}
\delta S_{1}=\dint_{0}^{\lambda _{1F}}[\frac{\partial L_{1}}{\partial 
\boldsymbol{x}_{1}}\cdot \boldsymbol{b}_{1}+\frac{\partial L_{1}}{\partial 
\boldsymbol{\dot{x}}_{1}}\cdot \boldsymbol{\dot{b}}_{1}]d\lambda _{1},
\label{Frechet}
\end{equation}%
Even though the functional is already Frech\'{e}t differentiable in $N^{(1)}(%
\boldsymbol{x}_{1})$, the electromagnetic equations require at least a $%
C^{2} $ orbit, as follows ;--- For a $C^{2}$ orbit $\boldsymbol{x}_{1}$ the
second term on the right-hand side of Eq. (\ref{Frechet}) can be further
integrated by parts using (\ref{endpoint}) to yield a term linear in $%
\boldsymbol{b}_{1} $, so that $\delta S_{1}$ becomes

\begin{equation}
\delta S_{1}=\dint \boldsymbol{G}_{1}\cdot \boldsymbol{b}_{1}d\lambda _{1},
\label{Euler-Lagrange}
\end{equation}%
with

\begin{equation}
\boldsymbol{G}_{1}\equiv -\frac{d}{d\lambda _{1}}(\frac{\partial L_{1}}{%
\partial \boldsymbol{\dot{x}}_{1}})+\frac{\partial L_{1}}{\partial 
\boldsymbol{x}_{1}},  \label{gradient1}
\end{equation}%
where $L_{1}$ is defined by Eq. (\ref{partialLagrangian}) and $\boldsymbol{G}%
_{1}\in \tciLaplace ^{4}$ is defined only along any $C^{2}$ orbit of the
natural neighborhood$\ N^{(2)}(\boldsymbol{x}_{1})$. Notice that Eq. (\ref%
{partialLagrangian}) is independent of the parametrization and the
expression of $L_{1}$ in terms of $\lambda _{1}$ is obtained simply by
replacing $t_{1}$ with $\lambda _{1}$ in Eq. (\ref{partialLagrangian}).
Expressing Eq. (\ref{partialLagrangian}) as a function of $\lambda _{1}$ and
evaluating $\boldsymbol{G}_{1}$ with Eq. (\ref{gradient1}) yields

\begin{eqnarray}
\boldsymbol{G}_{1} &=&\frac{d}{d\lambda _{1}}(m_{1}\frac{\boldsymbol{\dot{x}}%
_{1}}{\sqrt{\boldsymbol{\dot{x}}_{1}\cdot \boldsymbol{\dot{x}}_{1}}}-\frac{%
\boldsymbol{\dot{x}}_{2+}}{2||J_{\lambda _{2}}^{+}||}-\frac{\boldsymbol{\dot{%
x}}_{2-}}{2||J_{\lambda _{2}}^{-}||})  \label{evalgrad1} \\
&&+\frac{\partial }{\partial \boldsymbol{x}_{1}}(\frac{\boldsymbol{\dot{x}}%
_{1}\cdot \boldsymbol{\dot{x}}_{2+}}{2||J_{\lambda _{2}}^{+}||}+\frac{%
\boldsymbol{\dot{x}}_{1}\cdot \boldsymbol{\dot{x}}_{2-}}{2||J_{\lambda
_{2}}^{-}||}),  \notag
\end{eqnarray}%
where the dot over $\boldsymbol{x}_{i}$ denotes derivative respect to $%
\lambda _{i}$ for $i=1,2$ and $\lambda _{2}$ is the arbitrary parameter of
trajectory $2$. The condition for an extremum that follows from Eq. (\ref%
{Euler-Lagrange}) is $\boldsymbol{G}_{1}=0$ (plus the symmetric condition $%
\boldsymbol{G}_{2}=0$ obtained by varying trajectory $2$). Notice that $%
\boldsymbol{G}_{1}$ must be zero only in the \emph{open} interval $%
(0,\lambda _{1F})$ because the integrand of Eq. (\ref{Euler-Lagrange})
vanishes at the boundaries with $\boldsymbol{b}_{1}$. To pass from Eq. (\ref%
{Frechet}) to Eq. (\ref{Euler-Lagrange}) the vanishing perturbations at $%
O_{A}$\ and $L_{B}$ were enough to get rid of the boundary terms. The
perturbations of velocity and acceleration are arbitrary at $O_{A}$\ and $%
L_{B}$ because there is no prescribed orbit either before $O_{A}$ or after $%
L_{B}$, while the velocity and acceleration perturbations at $L^{-}$and $%
O^{+}$ must vanish for a $C^{2}$ match with the histories. The condition $%
\boldsymbol{G}_{1}=0$\ defined by Eq. (\ref{evalgrad1}) yields the
electromagnetic equations of motion with the Li\'{e}nard-Wierchert-Lorentz
force\cite{Fey-Whe}, as evaluated in the Appendix. The gradient for
variations of trajectory $2$\ is obtained analogously, by discarding the
integration over the past history of particle $2$\ and defining a
sub-functional $S_{2}(\boldsymbol{b}_{2},\boldsymbol{\dot{b}}_{2})$obtained
from the above $S_{1}$\ (\ref{partialLagrangian}) by exchanging particle
indices. The Banach space for arbitrary $C^{2}$ variations of both
trajectories respecting the EHBCs is the direct product $N^{(2)}(\boldsymbol{%
x}_{1})\otimes N^{(2)}(\boldsymbol{x}_{2})\equiv N^{(2)}(\boldsymbol{x}_{1,}%
\boldsymbol{x}_{2})$ with the norm given by $|\boldsymbol{b}_{1},\boldsymbol{%
b}_{2}|_{N(x_{1},x_{2})}\equiv \sup ||\boldsymbol{\dot{b}}_{1}||_{4}+\sup ||%
\boldsymbol{\dot{b}}_{2}||_{4}+\sup ||\boldsymbol{\ddot{b}}_{1}||_{4}+\sup ||%
\boldsymbol{\ddot{b}}_{2}||_{4}$, which is the natural physical space of
orbits satisfying the EHBCs.

Action (\ref{Darw}) is not defined for the superluminal elements of $N^{(1)}(%
\boldsymbol{x}_{1})$ (which have a large norm $|\boldsymbol{b}%
_{1}|_{N(x_{1})}$) because it involves taking the square-root of a negative
number. The above defined norm guarantees that sufficiently small
neighborhoods of \emph{subluminal} orbits contain only subluminal orbits (by 
\textit{Proposition 1}), but the set of subluminal orbits is not closed
because Cauchy sequences of subluminal orbits can converge to \emph{luminal}
orbits. Moreover, luminal orbits can have small neighborhoods containing
superluminal orbits, for which again action (\ref{Darw}) is not even
defined. In the following we relax the parametrization-invariance and
construct a second Poincar\`{e}-invariant functional defined everywhere in $%
N^{(1)}(\boldsymbol{x}_{1})$ and yielding the same electromagnetic equations
of motion. For superluminal trajectories the lightcone condition (\ref%
{implicit}) can have an arbitrary number of zeros, and for these the double
integration on the right-hand-side of Eq. (\ref{aFokker}) is generalized by
extending formula (\ref{limit}) to all zeros of (\ref{implicit}) in the
integration interval $(0,\lambda _{1F})$, which prescribes a vanishing
integral in the case of no solution in the interval. Our second functional
is obtained by further generalizing the kinetic terms, i.e.,%
\begin{eqnarray}
\Omega &\equiv &-\tint\limits_{0}^{\lambda _{1F}}\frac{m_{1}}{2p}(%
\boldsymbol{\dot{x}}_{1}\cdot \boldsymbol{\dot{x}}_{1})^{p}d\lambda
_{1}-\tint\limits_{0}^{\lambda _{2F}}\frac{m_{2}}{2p}(\boldsymbol{\dot{x}}%
_{2}\cdot \boldsymbol{\dot{x}}_{2})^{p}d\lambda _{2}  \notag \\
&&+\tint\limits_{0}^{\lambda _{jF}}\boldsymbol{A(\boldsymbol{x}_{j})\cdot 
\dot{x}}_{j}d\lambda _{j}.  \label{Second}
\end{eqnarray}%
The last term of action (\ref{Second}) is the double integral of action (\ref%
{Darw}) written in a convenient form and extended to arbitrary orbits by
evaluating $\boldsymbol{A(\boldsymbol{x}_{j})}$ with Eq.(\ref{defvecA})
extended to all the zeros of the lightcone condition inside $[0,\lambda
_{kF}]$. Notice that the last term of (\ref{Second}) is still
parametrization-independent, unlike the generalized kinetic terms of (\ref%
{Second}) that are parametrization-invariant only if $p=1/2$. The
Euler-Lagrange condition of extremum (\ref{gradient1}) applied to action (%
\ref{Second}) yields%
\begin{equation}
m_{i}\frac{d}{d\lambda _{i}}[(\boldsymbol{\dot{x}}_{i}\cdot \boldsymbol{\dot{%
x}}_{i})^{p-1}\boldsymbol{\dot{x}}_{i}^{\mu }]=\tsum\limits_{k=1}^{4}%
\boldsymbol{\dot{x}}_{i}^{k}(\partial _{ki}\boldsymbol{A}^{\mu }-\partial
_{\mu i}\boldsymbol{A}^{k}),  \label{antisymmetric}
\end{equation}%
where the partial derivative respect to the covariant components is defined
by $\partial _{ki}\equiv \partial /\partial x_{i}^{k}$ and we expressed the
Euler Lagrange condition (\ref{gradient1}) leaving the kinetic terms on the
right-hand side. Equation (\ref{antisymmetric}) involves an anti-symmetric
tensor on the left-hand side, so that the Minkowski scalar product of (\ref%
{antisymmetric}) with the four-velocity $\boldsymbol{\dot{x}}_{i}$ yields
zero on the left-hand side, i.e., 
\begin{equation}
\frac{(2p-1)}{2p}\frac{d}{d\lambda _{i}}[(\boldsymbol{\dot{x}}_{i}\cdot 
\boldsymbol{\dot{x}}_{i})^{p}]=0.  \label{separation}
\end{equation}%
The Fokker-like action (\ref{Darw}) has $p=1/2$ so that Eq. (\ref{separation}%
) holds trivially, but for $p\neq 1/2$ condition (\ref{separation}) implies
that $(\boldsymbol{\dot{x}}_{i}\cdot \boldsymbol{\dot{x}}_{i})$ must be
constant along the extremum orbit. Moreover, for $p\neq 1/2$ action (\ref%
{Second}) is no longer parameter independent, and Eq. (\ref{separation})
shows that the extremum condition of (\ref{Second}) is expressed in a
parameter that along subluminal orbits is proportional to the proper-time
parameter (the constant of proportionality renormalizes the scalar mass of
each particle). Property (\ref{separation}) divides the orbits in three
invariant classes, as follows (a) if condition (\ref{dtau}) is positive at
any orbital point, then it must be positive at all points of an extremal
orbit, and (b) if the particle ever travels \emph{at} the speed of light,
then $(\boldsymbol{\dot{x}}_{i}\cdot \boldsymbol{\dot{x}}_{i})=0$ everywhere
along the extremal orbit, so that the particle travels at the speed of light 
\emph{everywhere} and last (c) a superluminal orbit is superluminal
everywhere. By combining the three types of trajectory for each particle we
can produce six different classes of orbits, luminal $1$-luminal $2$,
superluminal $1$-luminal $2$, and etc\ldots We henceforth take $p=1$ so that
the kinetic integrand (\ref{Second}) is analytic and more important action (%
\ref{Second}) is defined everywhere and Frech\'{e}t-differentiable
everywhere in the Banach space $N^{(1)}(\boldsymbol{x}_{1})$. The advantages
of an action defined for trajectories violating (\ref{dtau}) is that the
ambient space of the functional (\ref{Second}) can be a complete normed
linear space, even if we later decide that only subluminal orbits are
interesting for physics.

\ 

\section{The second variation}

Here we calculate the second variation about the $C^{\infty }$%
low-velocity-circular-orbit-extrema $(\boldsymbol{x}_{1}^{c},\boldsymbol{x}%
_{2}^{c})$ of large enough radii \cite{Schoenberg,Schild} using either
action (\ref{Darw}) or (\ref{Second}). To calculate the first variation, in
Ref. \cite{stiff-hydrogen} we have expanded the delayed arguments of action (%
\ref{Darw}), a method that becomes cumbersome for the second variation. Here
we use a method motivated in the derivation of the low-velocity-limit of the
Fokker action \cite{Anderson}, only that ours is not restricted to
low-velocities and includes delay. Our method is equivalent to expanding the
delayed arguments of action (\ref{Darw}) but we use a shortcut equivalent to
taking derivatives of the Dirac delta-function, as done in Ref. \cite%
{Anderson}. We start with a definition for the derivative of the
right-hand-side of (\ref{limit}) and (\ref{limit1}). The following
proposition justifies the formal manipulation of the $\delta $ symbol inside
integration-by-parts formulas as long as the integrand vanishes at the
endpoints of the integration interval. To motivate our next definition we
start from formulas (\ref{limit}) and (\ref{limit1}) with trajectories given
by a perturbed circular orbit, i.e., $d(\lambda _{1},\lambda
_{2},\varepsilon )=d_{c}(\lambda _{1},\lambda _{2})+\varepsilon u(\lambda
_{1},\lambda _{2},\varepsilon )$ with $u(\lambda _{1},\lambda
_{2},\varepsilon )$ given by a polynomial function of the $\boldsymbol{b}%
_{1}(\lambda _{1})$ vanishing at the endpoints of $[L_{2I},L_{2F}]$
according to (\ref{endpoint}). Definition 1 yields 
\begin{eqnarray}
&&\dint\limits_{L_{2I}}^{L_{2F}}\delta (d(\lambda _{1},\lambda ,\varepsilon
))f(\lambda _{1},\lambda )d\lambda  \label{definieps} \\
&\equiv &\tsum\limits_{\bar{\lambda}_{2}^{(j)}}\frac{f(\lambda _{1},\bar{%
\lambda}_{2}^{(j)})}{||2J_{\lambda _{2}}||},  \notag
\end{eqnarray}%
where $2J_{\lambda _{2}}\equiv -\frac{\partial d}{\partial \lambda _{2}}%
(\lambda _{1},\bar{\lambda}_{2}^{(j)},\varepsilon )$ and the summation of
Eq. (\ref{definieps}) is extended to all zeros $(\lambda _{1},\bar{\lambda}%
_{2}^{(j)})$ of $d(\lambda _{1},\lambda _{2},\varepsilon )$ with $\bar{%
\lambda}_{2}^{(j)}\in $ $[L_{2I},L_{2F}]$ for any fixed $\lambda _{1}\in
\lbrack L_{1I},L_{1F}]$. The condition $u(\lambda _{1},\lambda
_{2},\varepsilon )=0$ at the endpoints ensures that the lightcone condition
is not perturbed at the endpoints, so that no zero $\bar{\lambda}_{2}^{(j)}$
of $d(\lambda _{1},\lambda ,\varepsilon )$ leaves or enters the interval $%
[L_{2I},L_{2F}]$ for small $\varepsilon $. The implicit function theorem for 
$d(\lambda _{1},\lambda ,\varepsilon )=0$ defines $\bar{\lambda}_{2}^{(j)}$
as a function of $\varepsilon $ with derivative%
\begin{equation}
\frac{\partial \bar{\lambda}_{2}^{(j)}}{\partial \varepsilon }=-\frac{\frac{%
\partial d}{\partial \varepsilon }(\lambda _{1},\bar{\lambda}%
_{2}^{(j)},\varepsilon )}{\frac{\partial d}{\partial \lambda _{2}}(\lambda
_{1},\bar{\lambda}_{2}^{(j)},\varepsilon )}.  \label{derivalamb}
\end{equation}%
The derivative of the right-hand-side of Eq. (\ref{definieps}) respect to $%
\varepsilon $ can be expressed with the help of (\ref{derivalamb}) in the
form%
\begin{equation}
\tsum\limits_{\bar{\lambda}_{2}^{(j)}}\frac{\partial _{\lambda
_{2}}[f(\lambda _{1},\lambda _{2})d_{\varepsilon }/2J_{\lambda _{2}}]}{%
||2J_{\lambda _{2}}(\lambda _{1},\lambda _{2})||}|_{\bar{\lambda}_{2}^{(j)}},
\label{derivative}
\end{equation}%
where $2J_{\lambda _{2}}\equiv -\frac{\partial d}{\partial \lambda _{2}}$
and $d_{\varepsilon }\equiv \frac{\partial d}{\partial \varepsilon }(\lambda
_{1},\lambda _{2},\varepsilon )$. Equation (\ref{derivative}) is formula (%
\ref{limit}) with $f(\lambda _{1},\lambda _{2})$ replaced by $\partial
_{\lambda }(f(\lambda _{1},\lambda _{2})d_{\varepsilon }/2J_{\lambda _{2}})$%
, an equality that justifies the use of a formal derivative of the
delta-function symbol as follows%
\begin{eqnarray}
&&\frac{d}{d\varepsilon }\dint\limits_{L_{2I}}^{L_{2F}}\delta (d(\lambda
_{1},\lambda ,\varepsilon ))f(\lambda _{1},\lambda )d\lambda
\label{operation} \\
&\equiv &\dint\limits_{L_{2I}}^{L_{2F}}d_{\varepsilon }\delta ^{^{\prime
}}(d(\lambda _{1},\lambda ,\varepsilon )f(\lambda _{1},\lambda )d\lambda 
\notag \\
&\equiv &\dint\limits_{L_{2I}}^{L_{2F}}\partial _{\lambda }(f(\lambda
_{1},\lambda _{2})d_{\varepsilon }/2J_{\lambda })\delta (d(\lambda
_{1},\lambda ,\varepsilon ))d\lambda  \notag
\end{eqnarray}%
where again $2J_{\lambda _{2}}\equiv -\frac{\partial d}{\partial \lambda _{2}%
}(\lambda _{1},\lambda _{2},\varepsilon )$ and $d_{\varepsilon }\equiv \frac{%
\partial d}{\partial \varepsilon }(\lambda _{1},\lambda _{2},\varepsilon )$
vanishes at the integration limits. We henceforth use (\ref{operation}) to
define the formal derivatives of the delta symbol, stressing that there is
no distributional limit involved but rather the above-defined operation.
Moreover, actions (\ref{Second}) and (\ref{Darw}) depend on a double
integral, i.e., either one of formulas (\ref{equivalent}). The derivative of
the interaction $I$ defined by Eq. (\ref{definieps}) with an $\varepsilon $%
-dependent $d(\lambda _{1},\lambda _{2},\varepsilon )$ is given by either
one of formulas%
\begin{eqnarray}
\frac{\partial I}{\partial \varepsilon } &=&  \label{dIde} \\
&&\tint\limits_{L_{2I}}^{L_{2F}}d\lambda _{2}\tsum\limits_{k}\frac{\partial
_{\lambda _{1}}[(\boldsymbol{\dot{x}}_{1}\cdot \boldsymbol{\dot{x}}%
_{2})d_{\varepsilon }/2J_{\lambda _{1}}]}{||2J_{\lambda _{1}}||}|\bar{\lambda%
}_{1}^{(k)}  \notag \\
&&\tint\limits_{L_{1I}}^{L_{1F}}d\lambda _{1}\tsum\limits_{j}\frac{\partial
_{\lambda _{2}}[(\boldsymbol{\dot{x}}_{1}\cdot \boldsymbol{\dot{x}}%
_{2})d_{\varepsilon }/2J_{\lambda _{2}}]}{||2J_{\lambda _{2}}||}|\bar{\lambda%
}_{2}^{(j)}  \notag
\end{eqnarray}%
with $2J_{\lambda _{k}}\equiv -\frac{\partial d}{\partial \lambda _{k}}%
(\lambda _{1},\lambda _{2},\varepsilon )$, as long as $d_{\varepsilon
}\equiv \frac{\partial d}{\partial \varepsilon }(\lambda _{1},\lambda
_{2},\varepsilon )=0$ at the integration limits. Otherwise we might have to
chose the line of Eq. (\ref{dIde}) for which $\frac{\partial d}{\partial
\varepsilon }(\lambda _{1},\lambda _{2},\varepsilon )$ vanishes at the
integration limits.

For the second variation we vary both trajectories simultaneously according
to%
\begin{eqnarray}
\boldsymbol{x}_{1} &=&\boldsymbol{x}_{1}^{c}+\varepsilon \boldsymbol{b}%
_{1},\qquad \boldsymbol{x}_{2}=\boldsymbol{x}_{2}^{c}+\varepsilon 
\boldsymbol{b}_{2},  \label{2-variation} \\
\boldsymbol{\dot{x}}_{1} &=&\boldsymbol{\dot{x}}_{1}^{c}+\varepsilon 
\boldsymbol{\dot{b}}_{1},\qquad \boldsymbol{\dot{x}}_{2}=\boldsymbol{\dot{x}}%
_{2}^{c}+\varepsilon \boldsymbol{\dot{b}}_{2},  \notag
\end{eqnarray}%
$\varepsilon \in \lbrack 0,1]$. We require vanishing perturbations at the
endpoints, 
\begin{eqnarray}
\boldsymbol{b}_{1}(O_{A}) &=&\boldsymbol{b}_{1}(L_{-})=0,  \label{endpoints}
\\
\boldsymbol{b}_{2}(O^{+}) &=&\boldsymbol{b}_{2}(L_{B})=0,  \notag
\end{eqnarray}%
\emph{and} vanishing velocity and acceleration perturbations on the \emph{%
history }side of each trajectory, i.e., at point $O^{+}$ of trajectory $2$
and at point $L^{-}$ of trajectory $1$, 
\begin{eqnarray}
\boldsymbol{\dot{b}}_{1}(L^{-}) &=&\boldsymbol{\ddot{b}}_{2}(L^{-})=0,
\label{oneside} \\
\boldsymbol{\dot{b}}_{2}(O^{+}) &=&\boldsymbol{\ddot{b}}_{2}(O^{+})=0, 
\notag
\end{eqnarray}%
so that the trajectories can be continued to a $C^{2}$ trajectory $%
\boldsymbol{b}_{1}=0$ on the boundary segment $(L^{-},L^{+})$ and $%
\boldsymbol{b}_{2}=0$ on the boundary segment $(O^{-},O^{+})$. The quadratic
integrand of the Taylor expansion involves products of variations at points
connected by the lightcone condition (rather than variations at the same
time as in the Kepler problem). We expand the action in a Taylor series up
to the second order in $\varepsilon $ by using a directional derivative
along the $C^{2}$ trajectory variation (\ref{2-variation}) of $N^{(2)}(%
\boldsymbol{x}_{1}^{c},\boldsymbol{x}_{2}^{c})$. Once the circular orbit is
an extremum, the first variation vanishes so that Taylor's theorem gives the
functional at $\varepsilon =1$ as a sum of its value at $\varepsilon =0$
plus the second-variation evaluated at some $\varepsilon \in \lbrack 0,1]$.

The second variation of the first term on the right-hand-side of Eq. (\ref%
{Darw}), representing the kinetic energy is%
\begin{equation}
\Delta ^{(2)}K_{1}=m_{1}\varepsilon ^{2}\dint\limits_{O_{A}}^{L^{+}}d\lambda
_{1}[\frac{(\boldsymbol{\dot{x}}_{1}^{c}\cdot \boldsymbol{\dot{b}}_{1})^{2}-%
\boldsymbol{\dot{x}}_{1}^{c2}\boldsymbol{\dot{b}}_{1}^{2}}{2(\boldsymbol{%
\dot{x}}_{1}^{c})^{3/2}}].  \label{Kinetic}
\end{equation}%
Formula (\ref{Kinetic}) is positive-definite, which is seen as follows;---If 
$\boldsymbol{\dot{b}}_{1}$is time-like, the positivity is given by the
reverse-Schwartz inequality of time-like vectors mentioned in the
introduction, while for a space-like $\boldsymbol{\dot{b}}_{1}$ Eq. (\ref%
{Kinetic}) is a sum of positive terms. Since the interaction integral is
naturally expressed as a double integral times the Dirac delta-function we
henceforth normalize all integrals to that form. To normalize Eq. (\ref%
{Kinetic}) we simply add a dummy integration over $d\lambda _{2}$ multiplied
by the integrating factor $2||J_{\lambda _{2}}||d\lambda _{2}$ and use that $%
\tint_{O_{-}}^{L_{B}}2||J_{\lambda _{2}}||\delta _{D}^{c}d\lambda _{2}=1$,
yielding

\begin{equation}
\Delta ^{(2)}K_{1}=m_{1}\varepsilon
^{2}\dint\limits_{O_{-}}^{L_{B}}\dint\limits_{O_{A}}^{L^{+}}d\lambda
_{1}d\lambda _{2}[\frac{(\boldsymbol{\dot{x}}_{1}^{c}\cdot \boldsymbol{\dot{b%
}}_{1})^{2}-\boldsymbol{\dot{x}}_{1}^{c2}\boldsymbol{\dot{b}}_{1}^{2}}{(%
\boldsymbol{\dot{x}}_{1}^{c})^{3/2}}]||J_{\lambda _{2}}||\delta _{D}^{c}.
\label{rational-K1}
\end{equation}%
The symbol $\delta _{D}^{c}$ is an abbreviation for $\delta (d(\lambda
_{1},\lambda _{2},\varepsilon ))$ as of definitions (\ref{limit}) and (\ref%
{limit1}) while upper index $c$ denotes the circular-orbit functions. Notice
that the low-velocity-limit of $||J_{\lambda _{2}}||$ in particle-time
parametrization is the spatial separation in light-cone, $r_{12}$, as
defined by Eq. (\ref{Jacobian_t}). To abbreviate notation we henceforth
indicate the double-integral over both circular orbits of any integrand $%
g(\lambda _{1},\lambda _{2},\varepsilon )$ times $\delta (d(\lambda
_{1},\lambda _{2},\varepsilon ))$ by $\tint_{C}g$ . For example the kinetic
term Eq. (\ref{rational-K1}) is abbreviated to 
\begin{equation}
\Delta ^{(2)}K_{1}=m_{1}\varepsilon ^{2}\int\limits_{C}[\frac{(\boldsymbol{%
\dot{x}}_{1}^{c}\cdot \boldsymbol{\dot{b}}_{1})^{2}-\boldsymbol{\dot{x}}%
_{1}^{c2}\boldsymbol{\dot{b}}_{1}^{2}}{(\boldsymbol{\dot{x}}_{1}^{c})^{3/2}}%
]||J_{\lambda _{2}}||.  \label{short-K1}
\end{equation}%
Next we calculate the second-variation of the interaction term by
substituting variation (\ref{2-variation}) into the integrand $I_{F}\equiv
\delta (|\boldsymbol{x}_{1}-\boldsymbol{x}_{2}|^{2})\boldsymbol{\dot{x}}%
_{1}\cdot \boldsymbol{\dot{x}}_{2}$ and expand in a Taylor series in $%
\varepsilon $ using the above define rules for the formal derivative. The
separation $d(\lambda _{1},\lambda _{2},\varepsilon )$ is perturbed along
variation (\ref{2-variation}) to 
\begin{eqnarray}
d(\lambda _{1},\lambda _{2},\varepsilon ) &=&|\boldsymbol{x}_{1}-\boldsymbol{%
x}_{2}|^{2}  \notag \\
&=&|\boldsymbol{x}_{1}^{c}-\boldsymbol{x}_{2}^{c}|^{2}+2\varepsilon (%
\boldsymbol{x}_{1}^{c}-\boldsymbol{x}_{2}^{c})\cdot (\boldsymbol{b}_{1}-%
\boldsymbol{b}_{2})  \notag \\
&&+\varepsilon ^{2}|\boldsymbol{b}_{1}-\boldsymbol{b}_{2}|^{2},
\label{varicone}
\end{eqnarray}%
so that the formal expansion of $\delta _{D}$ becomes%
\begin{eqnarray}
\delta _{D} &=&\delta _{D}(|\boldsymbol{x}_{1}^{c}-\boldsymbol{x}%
_{2}^{c}|^{2})  \notag \\
&&+[2(\boldsymbol{x}_{1}^{c}-\boldsymbol{x}_{2}^{c})\cdot (\boldsymbol{b}%
_{1}-\boldsymbol{b}_{2})+|\boldsymbol{b}_{1}-\boldsymbol{b}_{2}|^{2}]\delta
_{D}^{^{\prime }}  \notag \\
&&+2[\boldsymbol{x}_{12}^{c}\cdot (\boldsymbol{b}_{1}-\boldsymbol{b}%
_{2})]^{2}\delta _{D}^{^{^{\prime \prime }}}+O(3),  \label{expandelta}
\end{eqnarray}%
where $\boldsymbol{x}_{12}^{c}\equiv (\boldsymbol{x}_{1}^{c}-\boldsymbol{x}%
_{2}^{c})$. The bilinear product $\boldsymbol{\dot{x}}_{1}\cdot \boldsymbol{%
\dot{x}}_{2}$ is perturbed to 
\begin{equation}
\boldsymbol{\dot{x}}_{1}\cdot \boldsymbol{\dot{x}}_{2}=\boldsymbol{\dot{x}}%
_{1}^{c}\cdot \boldsymbol{\dot{x}}_{2}^{c}+2\varepsilon (\boldsymbol{\dot{x}}%
_{1}^{c}\cdot \boldsymbol{b}_{1}+\boldsymbol{\dot{x}}_{2}^{c}\cdot 
\boldsymbol{b}_{2})+\boldsymbol{\dot{b}}_{1}\cdot \boldsymbol{\dot{b}}_{2}
\label{varibilinear}
\end{equation}%
Henceforth one or two primes over $\delta _{D}$ denote respectively one or
two formal derivatives as defined by formulas (\ref{operation}) and (\ref%
{dIde}). The quadratic term of the Taylor expansion of $I_{F}\equiv \delta (|%
\boldsymbol{x}_{1}-\boldsymbol{x}_{2}|^{2})\boldsymbol{\dot{x}}_{1}\cdot 
\boldsymbol{\dot{x}}_{2}$ is obtained multiplying (\ref{expandelta}) by (\ref%
{varibilinear}) and collecting the second order terms, yielding 
\begin{eqnarray}
\Delta ^{(2)}I_{F} &=&\varepsilon ^{2}\boldsymbol{\dot{b}}_{1}\cdot 
\boldsymbol{\dot{b}}_{2}\delta _{D}+\varepsilon ^{2}\boldsymbol{\dot{x}}%
_{1}^{c}\cdot \boldsymbol{\dot{x}}_{2}^{c}|\boldsymbol{b}_{1}-\boldsymbol{b}%
_{2}|^{2}\delta _{D}^{^{\prime }}  \notag \\
&&+2\varepsilon ^{2}[\boldsymbol{x}_{12}^{c}\cdot (\boldsymbol{b}_{1}-%
\boldsymbol{b}_{2})]\Delta (\boldsymbol{\dot{x}}_{1}\cdot \boldsymbol{\dot{x}%
}_{2})\delta _{D}^{^{\prime }}  \notag \\
&&+2\varepsilon ^{2}\boldsymbol{\dot{x}}_{1}^{c}\cdot \boldsymbol{\dot{x}}%
_{2}^{c}[\boldsymbol{x}_{12}^{c}\cdot (\boldsymbol{b}_{1}-\boldsymbol{b}%
_{2})]^{2}\delta _{D}^{^{^{\prime \prime }}}.  \label{sec-IF}
\end{eqnarray}%
where $\Delta (\boldsymbol{\dot{x}}_{1}\cdot \boldsymbol{\dot{x}}_{2})\equiv
(\boldsymbol{\dot{x}}_{1}^{c}\cdot \boldsymbol{\dot{b}}_{2}+\boldsymbol{\dot{%
x}}_{2}^{c}\cdot \boldsymbol{\dot{b}}_{1})$ and $\boldsymbol{x}%
_{12}^{c}\equiv (\boldsymbol{x}_{1}^{c}-\boldsymbol{x}_{2}^{c})$. The first
term on the right-hand side of Eq. (\ref{sec-IF}) is already in the
normalized form of Eq. (\ref{rational-K1}). We henceforth drop the $%
\varepsilon ^{2}$ factor of the second-order expansion. The second term on
the right-hand side of Eq. (\ref{sec-IF}) must be split in three monomials, $%
\boldsymbol{\dot{x}}_{1}^{c}\cdot \boldsymbol{\dot{x}}_{2}^{c}(\boldsymbol{b}%
_{1}^{2}-2\boldsymbol{b}_{1}\cdot \boldsymbol{b}_{2}+\boldsymbol{b}%
_{2}^{2})\delta _{D}^{^{\prime }}$, and the formal integration by parts to
get rid of the $\delta _{D}^{^{\prime }}$ must treat each monomial
differently cause formula (\ref{operation}) needs a vanishing perturbation
at the endpoints;-- For example the monomial $\boldsymbol{\dot{x}}%
_{1}^{c}\cdot \boldsymbol{\dot{x}}_{2}^{c}\boldsymbol{b}_{1}^{2}\delta
_{D}^{^{\prime }}$ must be dealt with according to the first line of (\ref%
{dIde}), i.e.,%
\begin{eqnarray}
\dint\limits_{O_{A}}^{L^{+}}\boldsymbol{\dot{x}}_{1}^{c}\cdot \boldsymbol{%
\dot{x}}_{2}^{c}\boldsymbol{b}_{1}^{2}\delta _{D}^{^{\prime }}d\lambda _{1}
&=&  \label{parts-1} \\
&&-\dint\limits_{O_{A}}^{L^{+}}d\lambda _{1}\delta _{D}\frac{\partial }{%
\partial \lambda _{1}}(\frac{\boldsymbol{\dot{x}}_{1}^{c}\cdot \boldsymbol{%
\dot{x}}_{2}^{c}\boldsymbol{b}_{1}^{2}}{2||J_{\lambda _{1}}||}),  \notag
\end{eqnarray}%
since the integrand on the left-hand side of Eq. (\ref{parts-1}) vanishes
with $\boldsymbol{b}_{1}^{2}$ at $L^{+}$and $O_{A}$ (the EHBCs). Notice that
the monomial with $\boldsymbol{b}_{2}^{2}$ does not vanish at $L^{+}$and $%
O_{A}$. In that case the integration of choice would be over $d\lambda _{2}$%
. Using the above term-wise integration, the second term of the first line
on the right-hand side of Eq. (\ref{sec-IF}) yields%
\begin{eqnarray}
&&\int\limits_{C}\frac{\partial }{\partial \lambda _{1}}[\frac{(\boldsymbol{%
\dot{x}}_{1}^{c}\cdot \boldsymbol{\dot{x}}_{2}^{c})(\boldsymbol{b}_{1}\cdot 
\boldsymbol{b}_{2}-\boldsymbol{b}_{1}^{2})}{2||J_{\lambda _{1}}||}]
\label{short-secterm} \\
&&+\int\limits_{C}\frac{\partial }{\partial \lambda _{2}}[\frac{(\boldsymbol{%
\dot{x}}_{1}^{c}\cdot \boldsymbol{\dot{x}}_{2}^{c})(\boldsymbol{b}_{1}\cdot 
\boldsymbol{b}_{2}-\boldsymbol{b}_{2}^{2})}{2||J_{\lambda _{2}}||}].  \notag
\end{eqnarray}%
Next integrating by parts on the second line of the right-hand-side of Eq. (%
\ref{sec-IF}) yields%
\begin{eqnarray}
&&-\int\limits_{C}\frac{\partial }{\partial \lambda _{1}}[\frac{\Delta (%
\boldsymbol{\dot{x}}_{1}^{c}\cdot \boldsymbol{\dot{x}}_{2}^{c})(\boldsymbol{x%
}_{12}^{c}\cdot \boldsymbol{b}_{1})}{||J_{\lambda _{1}}||}]  \notag \\
&&+\int\limits_{C}\frac{\partial }{\partial \lambda _{2}}[\frac{\Delta (%
\boldsymbol{\dot{x}}_{1}^{c}\cdot \boldsymbol{\dot{x}}_{2}^{c})(\boldsymbol{x%
}_{12}\cdot \boldsymbol{b}_{2})}{||J_{\lambda _{2}}||}],  \label{short-line2}
\end{eqnarray}%
where again $\Delta (\boldsymbol{\dot{x}}_{1}\cdot \boldsymbol{\dot{x}}%
_{2})\equiv (\boldsymbol{\dot{x}}_{1}^{c}\cdot \boldsymbol{\dot{b}}_{2}+%
\boldsymbol{\dot{x}}_{2}^{c}\cdot \boldsymbol{\dot{b}}_{1})$ and $%
\boldsymbol{x}_{12}^{c}\equiv (\boldsymbol{x}_{1}^{c}-\boldsymbol{x}%
_{2}^{c}) $. Last, the third line of the right-hand-side of Eq. (\ref{sec-IF}%
) is transformed after \emph{two} integrations by parts into 
\begin{eqnarray}
&&\frac{1}{2}\int\limits_{C}\frac{\partial }{\partial \lambda _{1}}[\frac{1}{%
||J_{\lambda _{1}}||}\frac{\partial }{\partial \lambda _{1}}[\frac{%
\boldsymbol{\dot{x}}_{1}^{c}\cdot \boldsymbol{\dot{x}}_{2}^{c}(\boldsymbol{x}%
_{12}^{c}\cdot \boldsymbol{b}_{1})^{2}}{||J_{\lambda _{1}}||}]]
\label{short-line3} \\
&&+\frac{1}{2}\int\limits_{C}\frac{\partial }{\partial \lambda _{2}}[\frac{1%
}{||J_{\lambda _{2}}||}\frac{\partial }{\partial \lambda _{2}}[\frac{%
\boldsymbol{\dot{x}}_{1}^{c}\cdot \boldsymbol{\dot{x}}_{2}^{c}(\boldsymbol{x}%
_{12}^{c}\cdot \boldsymbol{b}_{2})^{2}}{||J_{\lambda _{2}}||}]]  \notag \\
&&-\int\limits_{C}\frac{\partial }{\partial \lambda _{1}}[\frac{1}{%
||J_{\lambda _{1}}||}\frac{\partial }{\partial \lambda _{2}}[\frac{%
\boldsymbol{\dot{x}}_{1}^{c}\cdot \boldsymbol{\dot{x}}_{2}^{c}(\boldsymbol{x}%
_{12}^{c}\cdot \boldsymbol{b}_{2})(\boldsymbol{x}_{12}^{c}\cdot \boldsymbol{b%
}_{1})}{||J_{\lambda _{2}}||}]].  \notag
\end{eqnarray}%
Henceforth we specify the circular orbit adopting particle-time
parametrization, i.e., $\boldsymbol{b}_{i}\equiv (0,\mathbf{b}_{i})$ and $%
\boldsymbol{\dot{b}}_{i}\equiv (0,\mathbf{\dot{b}}_{i})$ and $\boldsymbol{%
\dot{b}}_{i}\cdot \boldsymbol{\dot{b}}_{j}\equiv -\mathbf{\dot{b}}_{i}\cdot 
\mathbf{\dot{b}}_{j}$, where a dot between the vector parts henceforth
denotes Cartesian product. The velocities along a limiting circular orbit of
large radius are given by $\boldsymbol{\dot{x}}_{i}^{c}=(1,\vec{v}_{i}^{c})$
with 
\begin{eqnarray}
\vec{v}_{1}^{c} &=&\frac{m_{2}}{M\sqrt{r_{12}}}\hat{v}\boldsymbol{(}t_{1}%
\boldsymbol{),}  \label{Kepler} \\
\vec{v}_{2}^{c} &=&-\frac{m_{1}}{M\sqrt{r_{12}}}\hat{v}\boldsymbol{(}t_{2}%
\boldsymbol{).}  \notag
\end{eqnarray}%
In Eq. (\ref{Kepler}) $r_{12}$ is the constant separation in lightcone along
the circular orbit, $\hat{v}\boldsymbol{(}t\boldsymbol{)}$ is the unit
vector along the trajectory of particle $1$ and $M\equiv m_{1}+m_{2}$ (the
Kepler orbit is discussed in Ref. \cite{stiff-hydrogen} ). The period of the
circular orbit is given by Kepler's law 
\begin{equation}
T=2\pi \sqrt{\frac{M}{m_{1}m_{2}}}r_{12}^{3/2},  \label{period}
\end{equation}%
so that the lightcone separation $t_{1}=t_{2}\pm r_{12}$ is a negligible
fraction of the period for large $r_{12}$, i.e., the times in lightcone are
almost equal $(t_{1}\simeq t_{2})$, the spatial positions are almost in
diametral opposition and the velocities have nearly opposite directions.
Using the above circular orbit we calculate $||J_{t_{1}}||=||J_{t_{2}}||%
\simeq r_{12}$ and $\partial _{t_{1}}||J_{t_{1}}||=\partial
_{t_{2}}||J_{t_{2}}||=\boldsymbol{\dot{x}}_{1}^{c}\cdot \boldsymbol{\dot{x}}%
_{2}^{c}\simeq 1$ in the limit of a large $r_{12}$.

\textit{Theorem} : The second variation about circular orbits of large
enough radius is a strongly-positive quadratic form for $C^{2}$ trajectory
variations satisfying (\ref{endpoints}) and (\ref{oneside}).

Proof:---There are three basic types of integrals of quadratic monomials in
Eqs. (\ref{short-secterm}), (\ref{short-line2}) and (\ref{short-line3}),
namely (a) velocity-velocity, (b) position-position and (c)
position-velocity. Notice that integrals of type $\tint \delta _{D}(\mathbf{%
A\cdot \ddot{b}}_{i})(\mathbf{B\cdot b}_{j})$\ can be re-expressed as an
integral of a quadratic form of position and velocity variations only using (%
\ref{dIde}). In the following we inspect each type of integral, finding that
(a) is strongly-positive while (b) and (c) are dominated by (a) at large
enough radii, as follows;--

(a) The velocity-velocity terms of the second-variation are%
\begin{equation}
\Delta ^{(2)}V=\int\limits_{C}(m_{1}r_{12}\mathbf{\dot{b}}%
_{1}^{2}+m_{2}r_{21}\mathbf{\dot{b}}_{2}^{2}+\mathbf{\dot{b}}_{1}\cdot 
\mathbf{\dot{b}}_{2}),  \label{veve}
\end{equation}%
which is strongly positive-definite at large separations, $m_{i}r_{12}>>1$.
(b) The dominant quadratic terms in the displacements are%
\begin{eqnarray}
\Delta ^{(2)}R &=&\int\limits_{C}\frac{|\mathbf{b}_{1}-\mathbf{b}_{2}|^{2}}{%
2r_{12}^{2}}+\int\limits_{C}\frac{3(\hat{n}\cdot \mathbf{b}_{1}-\hat{n}\cdot 
\mathbf{b}_{2})^{2}}{2r_{12}^{2}}  \label{xx} \\
&&+\int\limits_{C}\frac{2(\hat{n}\cdot \mathbf{b}_{1})(\hat{n}\cdot \mathbf{b%
}_{2})}{r_{12}^{2}}  \notag \\
&&+\int\limits_{C}\frac{(\vec{v}_{2}\cdot \mathbf{b}_{1})(\vec{v}_{1}\cdot 
\mathbf{b}_{2})+(\vec{v}_{1}\cdot \mathbf{b}_{1})(\vec{v}_{2}\cdot \mathbf{b}%
_{2})}{r_{12}^{2}}.  \notag
\end{eqnarray}%
where $\hat{n}\equiv (\mathbf{b}_{1}-\mathbf{b}_{2})/r_{12}$. Quadratic form
(\ref{xx}) is \emph{not} positive-definite, and in fact for $\mathbf{b}_{1}=%
\mathbf{b}_{2}\equiv ||\boldsymbol{\delta }R||\hat{v}$ with $\hat{n}\cdot 
\hat{v}\mathbf{=0}$ we have $\Delta ^{(2)}R=-2m_{1}m_{2}||\boldsymbol{\delta 
}R||^{2}/(Mr_{12}^{3})$. The first two lines on the right-hand side of Eq. (%
\ref{xx}) have a non-negative sum, while we can show using (\ref{Kepler})
that the last line is bounded, i.e., 
\begin{equation}
\Delta ^{(2)}R\geq -\frac{m_{1}m_{2}}{Mr_{12}^{3}}\int\limits_{C}(||\mathbf{b%
}_{1}||^{2}+||\mathbf{b}_{2}||^{2}).  \label{boundR}
\end{equation}%
\textit{Lemma 1:}--For variations $\mathbf{b}_{i}$ vanishing at $t_{i}=0$
and $t_{i}=T_{\phi }$ it follows from the Fourier series $\mathbf{b}%
_{i}=\tsum \vec{a}_{k}\sin (\pi kt/T_{\phi })$ that%
\begin{equation}
\int\limits_{C}\boldsymbol{||}\mathbf{\dot{b}}_{i}||^{2}\geq \frac{\pi ^{2}}{%
T_{\phi }^{2}}\int\limits_{C}\boldsymbol{|}|\mathbf{b}_{i}||^{2},
\label{Fouri1}
\end{equation}%
where $\mathbf{\dot{b}}_{i}=d\mathbf{b}_{i}/dt_{i}$. In Eq. (\ref{Fouri1}), $%
T_{\phi }$ is the time for the circular rotation to travel the angle $\phi $
from $O_{A}$ to $L_{-}$, i.e., $T_{\phi }=(\phi /2\pi )T$ where $T$ \ is the
period as defined by Eq. (\ref{period}). The equal sign in (\ref{Fouri1})
holds \emph{iff }the first Fourier mode alone is present, i.e., $\vec{a}%
_{k}=0$ for $k\neq 1$. The following inequality is true for arbitrary arcs
of circle $\phi <2\pi $ but for simplicity we write it for EHBCs going a
complete turn, $T_{\phi }=T$, i.e., 
\begin{equation}
\int\limits_{C}r_{12}m_{i}\boldsymbol{||}\mathbf{\dot{b}}_{i}||^{2}\geq 
\frac{m_{1}m_{2}m_{i}}{4Mr_{12}^{2}}\int\limits_{C}\boldsymbol{|}|\mathbf{b}%
_{i}||^{2}.  \label{Fouri2}
\end{equation}%
Using Eqs. (\ref{Fouri2}) and (\ref{boundR}) we can show that an arbitrary
fraction $0<f<1$ of the kinetic term (\ref{veve}) dominates the quadratic
form (\ref{xx}) for sufficiently large $r_{12}$, i.e., $\ f\Delta
^{(2)}V\geq \Delta ^{(2)}R$.

(c) The quadratic terms involving position-velocity perturbations are also
dominated by the kinetic terms, as follows;--- Notice that the
position-velocity terms coming from (\ref{short-secterm}) integrate to zero,
i.e., 
\begin{eqnarray}
&&\frac{1}{2r_{12}}\int\limits_{C}(\mathbf{\dot{b}}_{1}\cdot \mathbf{b}_{2}-2%
\mathbf{\dot{b}}_{1}\cdot \mathbf{b}_{1})  \label{vanish} \\
&&+\frac{1}{2r_{12}}\int\limits_{C}(\mathbf{\dot{b}}_{2}\cdot \mathbf{b}%
_{1}-2\mathbf{\dot{b}}_{2}\cdot \mathbf{b}_{2}),  \notag
\end{eqnarray}%
where we used the large-radius limits $||J_{t_{1}}||=||J_{t_{2}}||\simeq
r_{12}$ and $\boldsymbol{\dot{x}}_{1}^{c}\cdot \boldsymbol{\dot{x}}%
_{2}^{c}\simeq 1$ and moved $r_{12}$ outside of the integration sign because
it is constant along circular orbits. After integration over one parameter
Eq. (\ref{vanish}) reduces to the integration of \ an exact differential
vanishing at the boundaries, so that (\ref{vanish}) vanishes. The largest
non-vanishing position-velocity terms come from (\ref{short-line2}) and (\ref%
{short-line3}), i.e.,

\begin{eqnarray}
\Delta ^{(2)}VR &=&  \label{VR} \\
&&+\int\limits_{C}\frac{1}{r_{12}}\Delta (\vec{v}_{1}\cdot \vec{v}_{2})(\hat{%
n}\cdot \mathbf{b}_{1}-\hat{n}\cdot \mathbf{b}_{2})  \notag \\
&&-\int\limits_{C}\frac{1}{r_{12}}\Delta (\vec{v}_{1}\cdot \vec{v}_{2})(\vec{%
v}_{1}\cdot \mathbf{b}_{1}-\vec{v}_{2}\cdot \mathbf{b}_{2}),  \notag
\end{eqnarray}%
where $\hat{n}\equiv (\mathbf{b}_{1}-\mathbf{b}_{2})/r_{12}$ and $\Delta (%
\vec{v}_{1}\cdot \vec{v}_{2})\equiv (\vec{v}_{1}^{c}\cdot \mathbf{b}_{2}+%
\vec{v}_{2}^{c}\cdot \mathbf{b}_{1})$. To show that the kinetic form (\ref%
{veve}) dominates the velocity-position quadratic terms for large enough $%
r_{12}$ we use inequality (\ref{Fouri2}) to derive \textit{Lemma 2}: 
\begin{eqnarray}
\int\limits_{C}r_{12}(m_{i}||\mathbf{\dot{b}}_{i}||^{2}+m_{j}||\mathbf{\dot{b%
}}_{j}||^{2}) &\geq &\int\limits_{C}(r_{12}m_{i}||\mathbf{\dot{b}}_{i}||^{2}
\label{Last} \\
&&+\int\limits_{C}\frac{m_{1}m_{2}m_{j}\boldsymbol{|}|\mathbf{b}_{j}||^{2}}{%
4Mr_{12}^{2}})  \notag \\
&\geq &\sqrt{\frac{m_{1}m_{2}m_{i}m_{j}}{Mr_{12}}}\int\limits_{C}||\mathbf{%
\dot{b}}_{i}||\boldsymbol{|}|\mathbf{b}_{j}||,  \notag
\end{eqnarray}%
where the last inequality is simply the completion of a binomial square. It
can be verified with Eq. (\ref{Kepler}) that the coefficients of the
monomials in the integrals of (\ref{VR}) are dominated by $1/r_{12}^{3/2}$,
so that \textit{Lemma 2} as of (\ref{Last}) is enough for the kinetic terms
to dominate all type (b) terms. To show that the second-variation is
positive-definite we divide the kinetic energy (\ref{veve}) in three equal
parts;--The first third dominates the position-squared terms (\ref{xx}) for
large enough $r_{12}$, as explained below Eq. (\ref{Fouri2}), while the
second third dominates the velocity-position terms by inequality (\ref{Last}%
). The last third is a non-degenerate positive-definite quadratic form of
the velocities, so that the second variation about circular orbits of large
enough radii is positive-definite, proving that circular orbits are local
minima. Moreover, the last third-part of Eq. (\ref{veve}) has all positive
eigenvalues, so that the second variation is \emph{strongly positive}.

\section{Conclusion and discussions}

An important question is the existence of an extremizing orbit for the
functional (\ref{Darw}) with arbitrary past data for particle $2$ plus
arbitrary future data for particle $1$, i.e., the existence result for
solutions of the mixed-type neutral-delay electromagnetic equations of
motion with general boundaries. There are no existence results for the
electromagnetic two-body problem apart from a few obtained for a
one-dimensional motion with repulsive interaction \cite{Driverfuture}, a
qualitatively different and simpler case where the equations are not neutral
but rather delay-only. For sufficiently small $C^{2}$ deformations of the
circular EHBCs preserving the boundary lightcones $O_{A}-O^{+}$, $%
O_{A}-O^{-} $ and $L^{-}-L_{B}$ and $L^{+}-L_{B}$, the second variation can
be proved positive-definite with analogous methods. \ Moreover, on a subset $%
\Theta \subset N^{(2)}(\boldsymbol{x}_{1}^{c},\boldsymbol{x}_{2}^{c})$ of
orbits satisfying $M\geq \sup (||\boldsymbol{\ddot{b}}_{1}||_{4})+\sup (||%
\boldsymbol{\ddot{b}}_{2}||_{4})$ for some $M$ we can reconstruct the $C^{2}$
perturbation using (\ref{endpoint}) and the one-sided conditions (\ref%
{oneside}), a formula analogous to Eq. (\ref{integral}). For example for $%
\boldsymbol{b}_{1}(\lambda _{1})$ we have%
\begin{equation}
\boldsymbol{b}_{1}(\lambda _{1})=\tint\limits_{\lambda _{c}}^{\lambda
_{1}}d\lambda _{b}\tint\limits_{\lambda _{1F}}^{\lambda _{c}}\boldsymbol{%
\ddot{b}}_{1}(\lambda _{a})d\lambda _{a},  \label{triple}
\end{equation}%
from which it follows that $\sup (||\boldsymbol{b}_{1}^{(k)}||)$ $\leq
\lambda _{1F}^{2-k}\sup ||\boldsymbol{\ddot{b}}_{1}||_{4}$, with an
analogous condition holding for $\boldsymbol{b}_{2}$ from the other side.
Conditions (\ref{oneside}) can be used to show that the $C^{2}$
perturbations inside $\Theta $ are equicontinuous and uniformly bounded, so
that by the Arzela-Ascoli theorem the set $\Theta $ is compact. If the
second variation is positive-definite, the functional is bounded from below
on the compact set $\Theta $ and assumes its minimum inside $\Theta $. We
conjecture that this point of minimum is an interior point of the compact
set. That granted, the minimum has a whole neighborhood inside $\Theta $, so
that Eq. (\ref{Euler-Lagrange}) holds for \emph{arbitrary} $\mathbf{b}_{k}$
and the gradients must vanish at the minimum. Conditions $G_{k}=0$ with $%
G_{k}$ defined by Eq. (\ref{evalgrad1}) are the state-dependent
neutral-delay equations of motion, so that this would be an existence result
for the state-dependent neutral-delay equations. This result would be the
analogous of the "Kurtzweil small delays don't matter theorem" for global
trajectories of DDE's on compact sets\cite{Kurtzweil}. The uniqueness theory
also differs from the case of Ref. \cite{wellposed}, and here one should
again start from the case of slightly perturbed circular boundaries, a case
where the equations of motion are approximated by neutral-delay-equations
with \emph{constant }advance and delay. For circular orbits of intermediate
radius some inspection suggests the minimum should become a saddle in a
bifurcation at a finite $O(1)$ radius in our unit system, i.e., of the order
of the classical electronic radius.

The existence proof \ is much harder for the solenoidal orbits discussed in
the appendix because of the denominators. For solenoidal orbits with a fast
velocity the functional might have a maximum as suggested by the kinetic
term.

A useful generalization of our second functional is for orbits defined on a
Sobolev space $H_{0}^{2}$ with derivatives defined almost everywhere. For
that we need to generalize the lightcone condition to arbitrary trajectories
and to generalize Eq. (\ref{limit}) to a sum over all zeros of the lightcone
condition. The fact that Eq. (\ref{limit}) is further integrated over the
other orbital parameter to make Eq. (\ref{equivalent}) compensates for the
extra zeros gained by changing the trajectories on a set of zero measure, so
that the functional can be defined on $H_{0}^{2}$. This generalization could
be useful in proving existence for the case of general boundaries.

Another\textbf{\ }question of interest regards the search for periodic
orbits and the possibility to restrict the variational method to the family
of periodic orbits satisfying the EHBCs. The reduction is possible to a
sub-family of periodic orbits by identifying the spatial components of $%
O_{A} $ with those of $L^{+}$ for trajectory $1$ and the spatial components
of $O_{-}$ with those of $L^{B}$ for trajectory $2$, which must be the case
along a periodic orbit. The orbital variation inside the family of periodic
orbits must preserve the history segment of each trajectory, as illustrated
in red in Fig. 3.1, which is a sub-family of the family of periodic orbits.
Last, it is possible to extremize the functionals directly in the space of $%
C^{1\text{ }}$orbits without even respecting the former sub-family
conditions. The conditions for an extremum with these most general
variations are no longer the electromagnetic equations of motion but rather
the overdetermined equations obtained by vanishing both linear terms on the
right-hand side of Eq. (\ref{Frechet}) \emph{separately}.

\bigskip

\section{Acknowledgements}

The author is solely responsible for errors even though there were useful
discussions with Savio Brochini Rodrigues, Hans-Otto Walther, Marcus V.
Lima, Tony Humphries, Michael Mackey, Tibor Krisztin, Clodoaldo Ragazzo,
Giorgio Fusco, Roger Nussbaum, John Mallet-Paret, and Nicola Guglielmi.

\textbf{\ }

\section{\protect\bigskip Appendix: Physics of the Fokker action}

Here we evaluate the gradient (\ref{evalgrad1}) explicitly and discuss the
physics of the two-body problem using proper-time parametrization for the
trajectories. The velocity respect to proper-time can be expressed either in
the form (\ref{physORB}) with $h_{1}=\gamma _{i}$, i.e.,

\begin{equation}
\mathbf{\upsilon }_{i}=(\gamma _{i},\gamma _{i}\vec{v}_{i}),
\label{Cartesianv}
\end{equation}%
or in the form%
\begin{equation}
\mathbf{\upsilon }_{i}=(\gamma _{i},\frac{d\vec{r}_{i}}{d\tau _{i}}),
\label{properv}
\end{equation}%
where%
\begin{equation}
\gamma _{i}\equiv \frac{dt_{i}}{d\tau _{i}}>0.  \label{gama}
\end{equation}%
According to Eq.(\ref{dtau}) the velocity respect to proper-time along
physical orbits, $\mathbf{\upsilon }_{i}\equiv d\boldsymbol{x}_{i}/d\tau
_{i} $, satisfies

\begin{equation}
(\mathbf{\upsilon }_{i}\cdot \mathbf{\upsilon }_{i})=1,  \label{constraint}
\end{equation}%
which can be solved for $\gamma _{i}$ using either Eq.(\ref{properv}) or Eq.(%
\ref{Cartesianv}), yielding%
\begin{equation}
\gamma _{i}=\sqrt{1+||\frac{d\vec{r}_{i}}{d\tau _{i}}||^{2}}=(1-||\vec{v}%
_{i}||^{2})^{-1/2}.  \label{gamav}
\end{equation}%
Notice that $||\frac{d\vec{r}_{i}}{d\tau _{i}}||$ is unbounded and becomes
arbitrarily large when the time-velocity approaches the speed of light.
Using condition (\ref{light-cone}) to solve for the Euclidean norm of the
spatial separation, we can express the separation vector $\boldsymbol{x}%
_{12}\equiv (\boldsymbol{x}_{1}-\boldsymbol{x}_{2})$ as 
\begin{equation}
\boldsymbol{x}_{12}=(\mp r_{12},r_{12}\hat{n}^{\pm }),  \label{normal}
\end{equation}%
where $r_{12\pm }$ is the distance in light-cone and $\mathbf{n}^{\pm }$ is
defined by%
\begin{equation}
\hat{n}^{\pm }\equiv \frac{\vec{r}_{1}(t_{1})-\vec{r}_{2}(t_{2\pm })}{%
r_{12\pm }},  \label{defn}
\end{equation}%
a unitary Euclidean three-vector. Notice that the spatial distance in
light-cone $r_{12\pm }\equiv ||\vec{r}_{2}(t_{2\pm })-\vec{r}_{1}(t_{1})||$
is a different function for each light-cone. The time-component of $\ 
\boldsymbol{x}_{12}\equiv (\boldsymbol{x}_{1}-\boldsymbol{x}_{2})$ is simply 
$t_{1}-t_{2}$ and evaluated with the negative sign of Eq.(\ref{retarded
light-cone}) (for the retarded cone) yields the positive number $%
t_{1}-t_{2-}=$ $r_{12-}\equiv ||\vec{r}_{2}(t_{2-})-\vec{r}_{1}(t_{1})||$.
The same Eq.(\ref{retarded light-cone}) with the plus sign (for the advanced
cone) yields the negative number $t_{1}-t_{2+}=$ $-r_{12+}\equiv -||\vec{r}%
_{2}(t_{2+})-\vec{r}_{1}(t_{1})||$. The sign of the time-component is
explicitly indicated by the plus or minus on the first entry of Eq.(\ref%
{normal}). \ The half-Jacobian (\ref{Jacobian}) with proper-time
parametrization is calculated using Eqs. (\ref{Cartesianv}) with index $1$
replaced by $2$ and definition (\ref{normal}), i.e.,%
\begin{equation}
J_{\tau _{2}}^{\pm }=(\boldsymbol{x}_{12}\cdot \boldsymbol{\dot{x}}_{2\pm
})=\mp \gamma _{2}r_{12\pm }(1\pm \hat{n}\cdot \vec{v}_{2})_{\pm },
\label{Jacotau}
\end{equation}%
where overdot represents derivative respect to proper-time of particle $2$.
Notice that $J_{\tau _{2}}^{\pm }$ can become singular when the particle
moves near the speed of light.

The partial derivative of $\tau _{2}^{\pm }$ with respect to $\mathbf{x}_{1}$
in Eq.(\ref{evalgrad1}) along the fixed trajectory of particle $2$ is given
by formula (\ref{slide}) with $\partial t_{2}^{\pm }$ replaced by $\partial
\tau _{2}^{\pm }$. \ Since we are operating with proper-time we can set $%
\sqrt{\boldsymbol{\dot{x}}_{1}\cdot \boldsymbol{\dot{x}}_{1}}=1$ in the
first denominator on the right-hand-side of Eq. (\ref{evalgrad1}), yielding 
\begin{eqnarray}
\boldsymbol{G}_{1} &=&m_{1}\boldsymbol{\ddot{x}}_{1}  \label{grad} \\
&&-\frac{d}{d\tau _{1}}(\frac{\boldsymbol{\dot{x}}_{2-}}{2J_{\tau _{2}}^{-}}-%
\frac{\boldsymbol{\dot{x}}_{2+}}{2J_{\tau _{2}}^{+}})  \notag \\
&&+\frac{\partial }{\partial \mathbf{x}_{1}}(\frac{\boldsymbol{\dot{x}}%
_{1}\cdot \boldsymbol{\dot{x}}_{2-}}{2J_{\tau _{2}}^{-}}-\frac{\boldsymbol{%
\dot{x}}_{1}\cdot \boldsymbol{\dot{x}}_{2+}}{2J_{\tau _{2}}^{+}}),  \notag
\end{eqnarray}%
where we took out the modulus sign using that $J_{\tau _{2}}^{+}$ is
negative and $J_{\tau _{2}}^{-}$ is positive. The derivative respect to $%
\tau _{1}$ on the right-hand side of Eq.(\ref{grad}) also acts on the
arguments $\tau _{2}^{\pm }$ since these are functions of $\tau _{1}$ by the
light-cone conditions%
\begin{equation}
|\boldsymbol{x}_{1}(\tau _{1})-\boldsymbol{x}_{2}(\tau _{2}^{\pm })|^{2}=0.
\label{light-conezero}
\end{equation}%
To evaluate the derivative of the retarded/advanced proper-time $\tau
_{2}^{\pm }$ with respect to $\tau _{1}$ we take the differential of the
light-cone condition (\ref{light-conezero}), i.e., 
\begin{equation}
2(\boldsymbol{x}_{12}\cdot \boldsymbol{\dot{x}}_{1})_{\pm }d\tau _{1}-2(%
\boldsymbol{x}_{12}\cdot \boldsymbol{\dot{x}}_{2})_{\pm }d\tau _{2}^{\pm }=0,
\label{diferential}
\end{equation}%
which yields%
\begin{equation}
\frac{d\tau _{2}^{\pm }}{d\tau _{1}}=\frac{(\boldsymbol{x}_{12}\cdot 
\boldsymbol{\dot{x}}_{1})_{\pm }}{(\boldsymbol{x}_{12}\cdot \boldsymbol{\dot{%
x}}_{2})_{\pm }},  \label{rate}
\end{equation}%
where $\boldsymbol{x}_{12\pm }\equiv (\boldsymbol{x}_{1}-\boldsymbol{x}%
_{2\pm })$. Formula (\ref{rate}) is valid for both the retarded and the
advanced lightcones. The same separation $\boldsymbol{x}_{12}$ appears on
both numerator and denominator on the right-hand-side of Eq.(\ref{rate}), so
that the plus or minus sign of Eq.(\ref{normal}) cancels out and the
derivative (\ref{rate}) is \emph{always positive} as it should be. We can
use the two signs of Eq.(\ref{rate}) to calculate the derivative of the most
retarded argument with respect to the most advanced argument by the chain
rule 
\begin{equation}
\frac{d\tau _{2-}}{d\tau _{2+}}=(\frac{d\tau _{2-}}{d\tau _{1}})(\frac{d\tau
_{1}}{d\tau _{2+}}),  \label{Tybor}
\end{equation}%
where $\frac{d\tau _{1}}{d\tau _{2_{+}}}\equiv (\frac{d\tau _{2+}}{d\tau _{1}%
})^{-1}$ and Eq.(\ref{Tybor}) is a non-negative rate because it is a product
of two positive factors. The fact that the retarded and the advanced
arguments have non-negative rates ensures the continuability of any
piecewise-continuous solution at a breaking point\cite{BellenZennaro}.
Therefore the usual mechanism for a neutral-delay equation to loose its
piece-wise continuous solution at a breaking point is absent and the neutral
equations of electrodynamics never loose solutions for this reason.

Using Eq.(\ref{rate}), the second line on the right-hand side of Eq.(\ref%
{grad}) evaluates to%
\begin{eqnarray}
&&\frac{1}{2(\boldsymbol{x}_{12}\cdot \boldsymbol{\dot{x}}_{2})_{-}^{2}}[%
\frac{d(\boldsymbol{x}_{12}\cdot \boldsymbol{\dot{x}}_{2})}{d\tau _{1}}%
\boldsymbol{\dot{x}}_{2}-(\boldsymbol{x}_{12}\cdot \boldsymbol{\dot{x}}_{1})%
\boldsymbol{a}_{2}]_{-}  \label{line1} \\
&&-\frac{1}{2(\boldsymbol{x}_{12}\cdot \boldsymbol{\dot{x}}_{2})_{+}^{2}}[%
\frac{d(\boldsymbol{x}_{12}\cdot \boldsymbol{\dot{x}}_{2})}{d\tau _{1}}%
\boldsymbol{\dot{x}}_{2}-(\boldsymbol{x}_{12}\cdot \boldsymbol{\dot{x}}_{1})%
\boldsymbol{a}_{2}]_{+}.  \notag
\end{eqnarray}%
where the lower index $\pm $ after the bracket indicates evaluation in the
advanced/retarded light-cone respectively and $\boldsymbol{a}_{2\pm }\equiv
d^{2}\boldsymbol{x}_{2}/d\tau _{2\pm }^{2}$ denotes the acceleration of
particle $2$ respect to proper time in the advanced/retarded light-cone
respectively. Last, on the third line of the right-hand side of Eq. (\ref%
{grad}) the partial derivative respect to $\mathbf{x}_{1}$ acts on $\mathbf{x%
}_{1}$ and also on quantities of particle $2$ by the rule%
\begin{equation}
\frac{\partial }{\partial \boldsymbol{x}_{1}}=\frac{\partial \tau _{2}}{%
\partial \boldsymbol{x}_{1}}\frac{d}{d\tau _{2}},  \label{E3}
\end{equation}%
with $\frac{\partial \tau _{2}}{\partial \boldsymbol{x}_{1}}$ given by Eq.(%
\ref{slide}). The manipulations are simple and the third line on the
right-hand side of Eq.(\ref{grad}) becomes%
\begin{eqnarray}
&&\frac{(\boldsymbol{\dot{x}}_{1}\cdot \boldsymbol{a}_{2})_{-}\boldsymbol{x}%
_{12-}}{2(\boldsymbol{x}_{12}\cdot \boldsymbol{\dot{x}}_{2})_{-}^{2}}-\frac{(%
\boldsymbol{\dot{x}}_{1}\cdot \boldsymbol{\dot{x}}_{2})_{-}\boldsymbol{\dot{x%
}}_{2-}}{2(\boldsymbol{x}_{12}\cdot \boldsymbol{\dot{x}}_{2})_{-}^{2}} 
\notag \\
&&+\frac{(\boldsymbol{\dot{x}}_{1}\cdot \boldsymbol{\dot{x}}_{2})_{-}[1-(%
\boldsymbol{a}_{2}\cdot \boldsymbol{x}_{12})_{-}]\boldsymbol{x}_{12-}}{2(%
\boldsymbol{x}_{12}\cdot \boldsymbol{\dot{x}}_{2})_{-}^{3}}
\label{line-last} \\
&&-\frac{(\boldsymbol{\dot{x}}_{1}\cdot \boldsymbol{a}_{2})_{+}\boldsymbol{x}%
_{12+}}{2(\boldsymbol{x}_{12}\cdot \boldsymbol{\dot{x}}_{2})_{+}^{2}}+\frac{(%
\boldsymbol{\dot{x}}_{1}\cdot \boldsymbol{\dot{x}}_{2})_{+}\boldsymbol{\dot{x%
}}_{2+}}{2(\boldsymbol{x}_{12}\cdot \boldsymbol{\dot{x}}_{2})_{+}^{2}} 
\notag \\
&&-\frac{(\boldsymbol{\dot{x}}_{1}\cdot \boldsymbol{\dot{x}}_{2})_{+}[1-(%
\boldsymbol{a}_{2}\cdot \boldsymbol{x}_{12})_{+}]\boldsymbol{x}_{12+}}{2(%
\boldsymbol{x}_{12}\cdot \boldsymbol{\dot{x}}_{2})_{+}^{3}}
\end{eqnarray}%
Using Eqs.(\ref{grad}), (\ref{line1}) and (\ref{line-last}) we can express
the gradient as 
\begin{equation}
\boldsymbol{G}_{1}=m_{1}\boldsymbol{\ddot{x}}_{1}-\frac{1}{2}F_{2}^{+}-\frac{%
1}{2}F_{2}^{-},  \label{physicalEq}
\end{equation}%
where 
\begin{eqnarray}
F_{2}^{\pm } &\equiv &\frac{(\boldsymbol{x}_{12}\cdot \mathbf{\upsilon }%
_{2})_{\pm }}{2\rho _{2}^{\pm }}[(\boldsymbol{x}_{12}\cdot \mathbf{\upsilon }%
_{1})\boldsymbol{a}_{2}-(\mathbf{\upsilon }_{1}\cdot \boldsymbol{a}_{2})%
\boldsymbol{x}_{12}]_{\pm }  \label{defA} \\
&&+\frac{(1-\boldsymbol{x}_{12}\cdot \boldsymbol{a}_{2})_{\pm }}{2\rho
_{2}^{\pm }}[(\boldsymbol{x}_{12}\cdot \mathbf{\upsilon }_{1})\mathbf{%
\upsilon }_{2}-(\mathbf{\upsilon }_{1}\cdot \mathbf{\upsilon }_{2})%
\boldsymbol{x}_{12}]_{\pm },  \notag
\end{eqnarray}%
where$\rho _{2}^{\pm }\equiv ||J_{\tau _{2}}^{\pm }||^{3}$ and $\mathbf{%
\upsilon }_{2\pm }\equiv d\boldsymbol{x}_{2}/d\tau _{2\pm }$.

The condition $\boldsymbol{G}_{1}=0$ yields a familiar Newtonian-like
equation of motion with the Lorentz-force of the other particle as a
semi-sum of advanced/retarded Li\'{e}nard-Wiechert fields, i.e., 
\begin{eqnarray}
m_{1}\frac{d\boldsymbol{\upsilon }_{1}}{d\tau _{1}} &=&\frac{1}{2}F_{2}^{-}(%
\boldsymbol{x}_{1},\mathbf{\upsilon }_{1},\boldsymbol{x}_{2}(\tau _{2}^{-}),%
\boldsymbol{\upsilon }_{2}(\tau _{2}^{-}),\mathbf{a}_{2}(\tau _{2}^{-}))
\label{Eq-electro} \\
&&+\frac{1}{2}F_{2}^{+}(\boldsymbol{x}_{1},\mathbf{\upsilon }_{1},%
\boldsymbol{x}_{2}(\tau _{2}^{+}),\boldsymbol{\upsilon }_{2}(\tau _{2}^{+}),%
\mathbf{a}_{2}(\tau _{2}^{+})),  \notag \\
m_{2}\frac{d\boldsymbol{\upsilon }_{2}}{d\tau _{2}} &=&\frac{1}{2}F_{1}^{-}(%
\boldsymbol{x}_{2},\mathbf{\upsilon }_{2},\boldsymbol{x}_{1}(\tau _{1}^{-}),%
\boldsymbol{\upsilon }_{1}(\tau _{1}^{-}),\mathbf{a}_{1}(\tau _{1}^{-}))
\label{Eq-proton} \\
&&+\frac{1}{2}F_{1}^{+}(\boldsymbol{x}_{2},\mathbf{\upsilon }_{2},%
\boldsymbol{x}_{1}(\tau _{1}^{+}),\boldsymbol{\upsilon }_{1}(\tau _{1}^{+}),%
\mathbf{a}_{1}(\tau _{1}^{+})),  \notag
\end{eqnarray}%
where $\boldsymbol{x}_{i},\boldsymbol{\upsilon }_{i},\boldsymbol{a}_{i}$\
are respectively the position, velocity and acceleration of particle $i$
with respect to proper-time $\tau _{i}$. In Eqs. (\ref{Eq-electro}) and (\ref%
{Eq-proton}) the forces $F_{k}^{\pm }$ depend respectively on the other
particle%
\'{}%
s retarded/advanced position, velocity and acceleration, as well as on the
object-particle`s present position and velocity. Moreover the
retarded/advanced points are implicitly defined by Eq. (\ref{retarded
light-cone}), so that Eqs. (\ref{Eq-electro}) and (\ref{Eq-proton}) are
neutral-delay equations of mixed-type with implicit state-dependent delay.
The forces $F_{k}^{\pm }$ of Eqs. (\ref{Eq-electro}) and (\ref{Eq-proton})
are the Lorentz force of the Li\'{e}nard-Wierchert fields of standard
electrodynamics textbooks\cite{Jackson,Barut}. Notice that each line of Eq. (%
\ref{defA}) is orthogonal to $\mathbf{\upsilon }_{1}$, so that it follows
from Eq. (\ref{Eq-electro}) that%
\begin{equation}
(\mathbf{\upsilon }_{1}\cdot \mathbf{\dot{\upsilon}}_{1})=0,
\label{orthogonal}
\end{equation}%
in agreement with Eq. (\ref{separation}) for $p\neq 1/2$ and for $p=1/2$ Eq.(%
\ref{orthogonal}) is the definition of proper-time parametrization.
Condition (\ref{constraint}) can be solved for the time-velocity with Eq. (%
\ref{gamav}), thereby reducing the dynamics to the spatial components of
Eqs. (\ref{Eq-electro}) and (\ref{Eq-proton}). Last, in the following we
discuss the denominators of Eqs. (\ref{Eq-electro}) and (\ref{Eq-proton}).\
The Li\'{e}nard-Wierchert force (\ref{defA}) involves denominators of type (%
\ref{Jacotau}), which become singular when the other particle travels near
the speed of light. This is illustrated expressing the vector-part of the
equation of motion (\ref{Eq-electro}) using particle-time parametrization
and using only the leading singular term of force (\ref{defA}), i.e., 
\begin{equation}
\frac{d}{dt}(\frac{m_{i}\vec{v}_{i}}{\sqrt{1-|\vec{v}_{i}|^{2}}})=\frac{-1}{%
r_{12}(1\pm \hat{n}\cdot \vec{v}_{i})^{3}}\hat{n}\times \mathbf{(}\hat{n}%
\times \vec{a}_{j})+...  \label{eqmot}
\end{equation}%
Notice that the left-hand-side of Eq. (\ref{eqmot}) becomes singular when 
\emph{particle }$i$ travels near the speed of light while the right-hand
side of Eq. (\ref{eqmot}) becomes singular when \emph{particle }$j$ travels
near the speed of light in either the past/future lightcone points, so that
if the two motions synchronize a solenoidal orbit with a fast velocity could
exist, as suggested in Ref. \cite{stiff-hydrogen}. Surprisingly this
non-trivial motion does not require large total momenta, as follows;---
Action (\ref{Darw}) is invariant by the Lorentz group if one also moves the
boundary-condition-segments with the group element (the red segments of Fig.
3.1). Noether's theorem \cite{Anderson} applies to action (\ref{Darw}) in a
way completely analogous to the formal derivation of Schild\cite{Schild}, as
explained in Ref. \cite{Anderson}, yielding invariants defined by \emph{%
finite} integrals, i.e., 
\begin{eqnarray}
\mathbf{p}^{\mu } &=&\mathbf{p}_{1}^{\mu }(\tau _{1})+\mathbf{p}_{2}^{\mu
}(\tau _{2})+  \label{p1p2} \\
\! &&-2\dint\limits_{\tau _{1}}^{L^{+}}\dint\limits_{O^{-}}^{\tau _{2}}(%
\boldsymbol{x}_{1}^{\mu }-\boldsymbol{x}_{2}^{\mu })\delta ^{\prime }(|%
\boldsymbol{x}_{1}-\boldsymbol{x}_{2}|^{2})\boldsymbol{\dot{x}}_{1}\cdot 
\boldsymbol{\dot{x}}_{2}d\tau _{1}d\tau _{2}  \notag \\
&&\!+2\dint\limits_{O_{A}}^{\tau _{1}}\dint\limits_{\tau _{2}}^{L_{B}}(%
\boldsymbol{x}_{1}^{\mu }-\boldsymbol{x}_{2}^{\mu })\delta ^{\prime }(|%
\boldsymbol{x}_{1}-\boldsymbol{x}_{2}|^{2})\boldsymbol{\dot{x}}_{1}\cdot 
\boldsymbol{\dot{x}}_{2}d\tau _{1}d\tau _{2}.  \notag
\end{eqnarray}%
and%
\begin{eqnarray}
L^{\alpha \beta } &=&(\mathbf{r}_{1}^{\alpha }\mathbf{p}_{1}^{\beta }-%
\mathbf{r}_{1}^{\beta }\mathbf{p}_{1}^{\alpha })|_{\tau _{1}}+(\mathbf{r}%
_{2}^{\alpha }\mathbf{p}_{2}^{\beta }-\mathbf{r}_{2}^{\beta }\mathbf{p}%
_{2}^{\alpha })|_{\tau _{2}}  \notag \\
&&-2\dint\limits_{\tau _{1}}^{L^{+}}\dint\limits_{O^{-}}^{\tau _{2}}(%
\boldsymbol{x}_{1}^{\alpha }\boldsymbol{x}_{2}^{\beta }-\boldsymbol{x}%
_{1}^{\beta }\boldsymbol{x}_{2}^{\alpha })\delta ^{\prime }(|\boldsymbol{x}%
_{1}-\boldsymbol{x}_{2}|^{2})\boldsymbol{\dot{x}}_{1}\cdot \boldsymbol{\dot{x%
}}_{2}d\tau _{1}d\tau _{2}  \notag \\
&&+2\dint\limits_{O_{A}}^{\tau _{1}}\dint\limits_{\tau _{2}}^{L_{B}}(%
\boldsymbol{x}_{1}^{\alpha }\boldsymbol{x}_{2}^{\beta }-\boldsymbol{x}%
_{1}^{\beta }\boldsymbol{x}_{2}^{\alpha })\delta ^{\prime }(|\boldsymbol{x}%
_{1}-\boldsymbol{x}_{2}|^{2})\boldsymbol{\dot{x}}_{1}\cdot \boldsymbol{\dot{x%
}}_{2}d\tau _{1}d\tau _{2}  \notag \\
&&+\dint\limits_{\tau _{1}}^{L^{+}}\dint\limits_{O^{-}}^{\tau _{2}}(%
\boldsymbol{\dot{x}}_{1}^{\alpha }\boldsymbol{\dot{x}}_{2}^{\beta }-%
\boldsymbol{\dot{x}}_{1}^{\beta }\boldsymbol{\dot{x}}_{2}^{\alpha })\delta
^{\prime }(|\boldsymbol{x}_{1}-\boldsymbol{x}_{2}|^{2})d\tau _{1}d\tau _{2} 
\notag \\
&&-\dint\limits_{O_{A}}^{\tau _{1}}\dint\limits_{\tau _{2}}^{L_{B}}(%
\boldsymbol{\dot{x}}_{1}^{\alpha }\boldsymbol{\dot{x}}_{2}^{\beta }-%
\boldsymbol{\dot{x}}_{1}^{\beta }\boldsymbol{\dot{x}}_{2}^{\alpha })\delta
^{\prime }(|\boldsymbol{x}_{1}-\boldsymbol{x}_{2}|^{2})d\tau _{1}d\tau _{2}.
\label{angularT}
\end{eqnarray}

\bigskip

where 
\begin{eqnarray}
\mathbf{p}_{1}^{\mu } &\equiv &\frac{m_{1}\boldsymbol{v}_{1}^{\mu }}{\sqrt{1-%
\boldsymbol{|}\vec{v}_{1}\mathbf{|}^{2}}}  \label{p1} \\
&&-\frac{\boldsymbol{v}_{2-}^{\mu }}{2r_{12}^{-}(1-\hat{n}\cdot \vec{v}_{2-})%
}-\frac{\boldsymbol{v}_{2+}^{\mu }}{2r_{12}^{+}(1-\hat{n}\cdot \vec{v}_{2+})}%
,  \notag \\
\mathbf{p}_{2}^{\mu } &\equiv &\frac{m_{2}\boldsymbol{v}_{1}^{\mu }}{\sqrt{%
1-|\vec{v}_{2}\mathbf{|}^{2}}}  \label{p2} \\
&&-\frac{\boldsymbol{v}_{1-}^{\mu }}{2r_{21}^{-}(1-\hat{n}\cdot \vec{v}_{1-})%
}-\frac{\boldsymbol{v}_{1+}^{\mu }}{2r_{21}^{+}(1-\hat{n}\cdot \vec{v}_{1+})}%
.  \notag
\end{eqnarray}%
Notice that $\mathbf{p}_{1}$ and $\mathbf{p}_{2}$ as defined by Eqs. (\ref%
{p1}) and (\ref{p2}) can be small even at fast velocities, so that a
solenoidal motion with a stiff gyration near the speed of light is possible
with finite and small mechanical momenta (\ref{p1p2}), as illustrated in
Fig. 6.1. The solenoidal orbits of Ref. \cite{stiff-hydrogen} were estimated
to have a velocity near the speed of light, and the physical interest stems
from the fact that these can be found in the physical region of small
4-momentum and angular-momentum. At present there is no numerical integrator
available to integrate such non-trivial \ state-dependent neutral-delay
equations, and we hope this work contributes to the construction of such
integrator.


\begin{references}


\bibitem{Schwarz} K. Schwarzschild {\it Gottinger Nachrichten} {\bf 128}, 132 (1903).

\bibitem{Tetrode-Fokker} H. Tetrode {\it Z. Phys.} {\bf 10}, 317 (1922);
 A. D. Fokker {\it Z. Phys.} {\bf 58},  386 (1929),  A. D. Fokker {\it Physica } {\bf 9},  33 (1929), 
A. D. Fokker {\it Physica } {\bf 12},  145 (1932).

\bibitem{Fey-Whe} J. A. Wheeler and R. P. Feynman {\it Rev.Mod. Phys.} {\bf 17}, 
157 (1945);  J. A. Wheeler and R. P. Feynman {\bf 21}, 425 (1949). 

\bibitem{Marsden}J. Marsden and M. West, {\it Acta Numerica }, 357 (2001).

\bibitem{Leiter} D. Leiter {\it Am. J. Phys} {\bf 38}, 207 (1970). 

\bibitem{Jackson}J.D. Jackson, {\em Classical Electrodynamics} Second Edition, 
John Wiley and Sons, New York(1975).

\bibitem{EliezerReview}C. Jayaratnam Eliezer, {\it Reviews of Modern Physics} {\bf 19} (1947). 

\bibitem{Dirac} P. A. M.Dirac, {\it Proceedings of the Royal Society of London, ser. A}
{\bf 167},148 (1938).

\bibitem{Schoenberg}M. Schoenberg, {\it Physical Review} {\bf 69} 211 (1946).

\bibitem{Schild}A. Schild, {\it Physical Review} {\bf 131}, 2762 (1963).


\bibitem{stiff-hydrogen} J. De Luca, { \it Physical Review E } {\bf 73}, 026221 (2006).


\bibitem{Gordon}W. B. Gordon, {\it American Journal of Mathematics}, {\bf 99} 961 (1977).

\bibitem{Anderson} J.L. Anderson  {\em Principles of Relativity Physics },
Academic press, New York, (1967), page 225.

\bibitem{Narlikar}F. Hoyle and J. V. Narlikar, {\em Lectures on Cosmology and Action at a Distance Electrodynamics }, 
World Scientific, London (1996).

\bibitem{StarusCritique}A. Staruszkiewicz, {\it Annalen der Physik } {\bf 25} 362 (1970).


\bibitem{Kurtzweil} J. Kurtzweil, in {\em Lecture Notes in Mathematics -- Seminar on Differential Equations 
and Dynamical Systems, II } {\bf144}, 134 (1970), Edited by J.A. Yorke, Springer-Verlag, NY (1970).

\bibitem{Driverfuture}R.Driver, { \it Physical Review D } {\bf 19}, 1098 (1979).


\bibitem{neutralDriver}R. D. Driver, {\it J. Differential Equations} {\bf 54} 73 (1984).

\bibitem{MixedDriver}R.D. Driver, {\it Nonlinear Analysis} {\bf 8}, 155 (1984).

\bibitem{JMP}J. De Luca, {\it J. Math. Phys.} {\bf 48}, 012702  (2007).


\bibitem{wellposed}J. A. Murdock {\it Annals
of Physics} {\bf 84}, 432 (1974). 


\bibitem{BellenZennaro} A. Bellen and M. Zennaro, {\em Numerical Methods for Delay Differential Equations} 
Oxford University Press, NY (2003).

\bibitem{Barut} A. O. Barut, {\em Electrodynamics and Classical Theory of Fields and Particles} 
Dover, New York (1980).

\bibitem{Darwin} C.G.Darwin, {\em Philos. Mag. } {\bf 30}, 537 (1920).


\bibitem{Efy1}E.B.Hollander and J. De Luca, {\it Phys. Rev. E} {\bf 67 } 026219 (2003).


\bibitem{Efy2}E.B.Hollander and J. De Luca, {\it Chaos} {\bf 14 } 1093 (2004).


\bibitem{PRL} J. De Luca, {\it Phys. Rev. Lett.} {\bf 80}, 680 (1998),  
J. De Luca {\it Phys. Rev. E} {\bf 58}, 5727 (1998).


\bibitem{Simple}J. De Luca, {\it Phys. Rev. E} {\bf 62}, 2060 (2000).

\end{references}
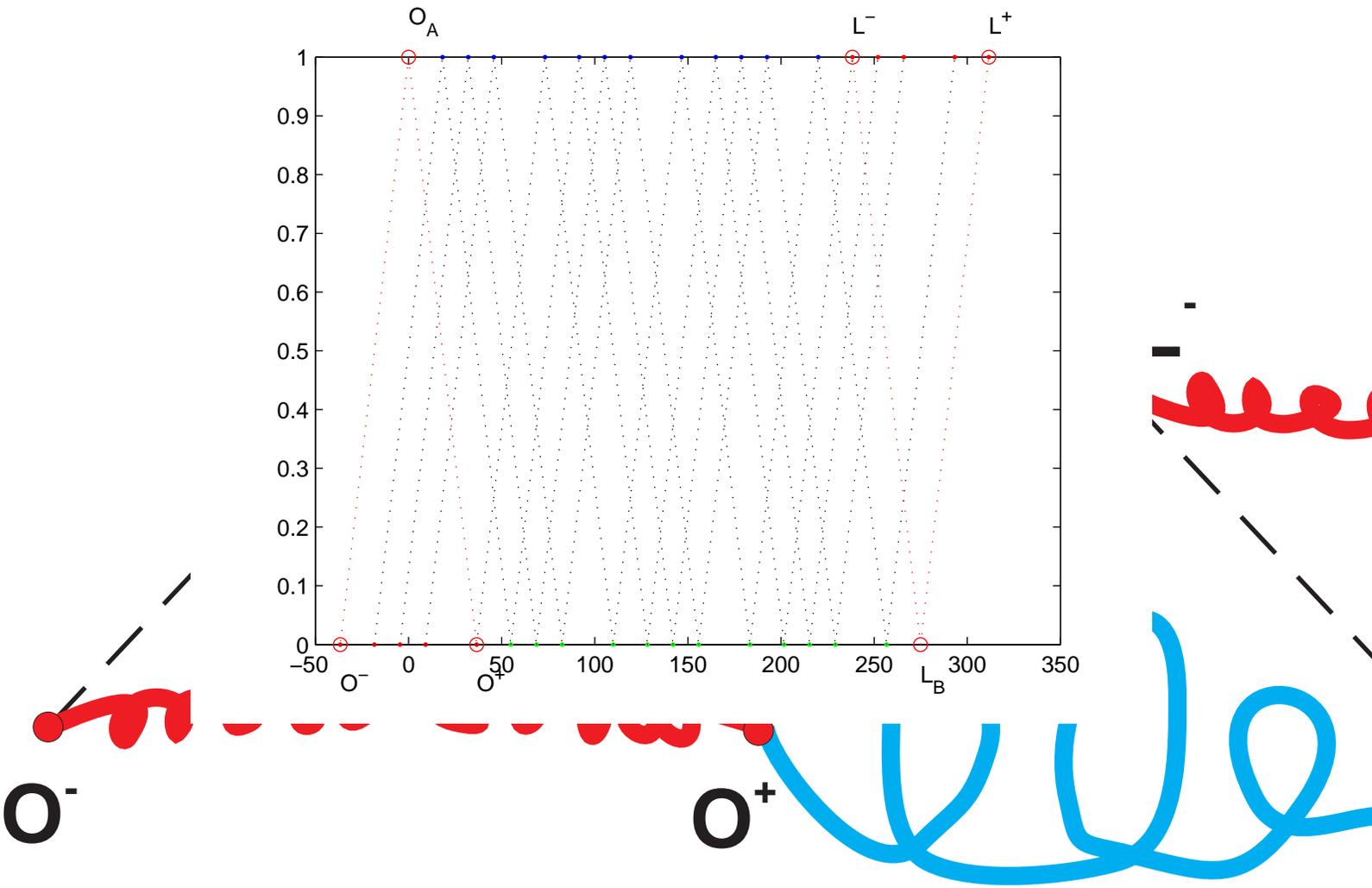
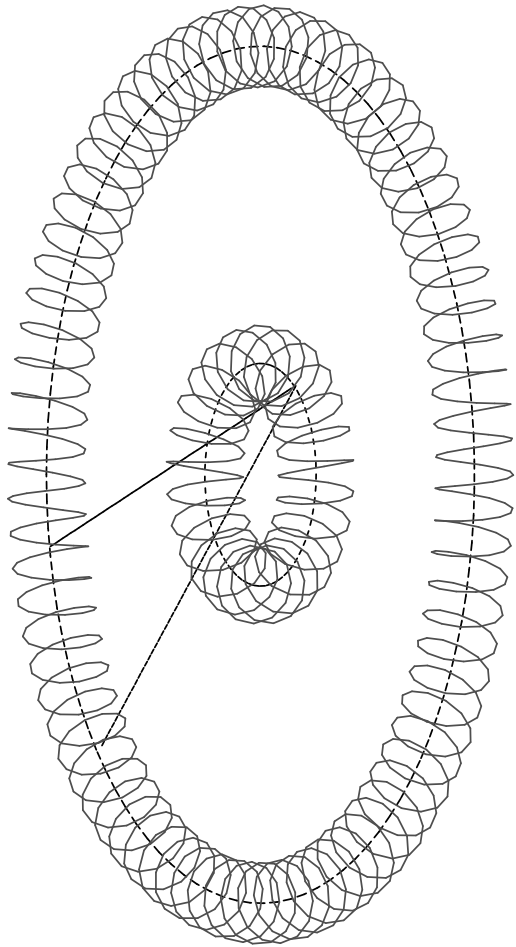
\end{document}